\documentclass[pre,showpacs,showkeys,preprintnumbers,amsmath,amssymb,superscriptaddress,twocolumn]{revtex4}
\usepackage{graphicx}
\usepackage{amsfonts}
\usepackage{curves}
\usepackage{amstext}

\def\E{{\mathbb E}}
\def\P{{\mathbb P}}
\def\R{{\mathbb R}}
\def\Z{{\mathbb Z}}

\def\N{{\mathcal N}}

\def\pa{\partial\Omega}

\def\ve{\varepsilon}

\begin{document}

\title{A Multiscale Guide to Brownian Motion}
\author{Denis S. Grebenkov}
 \email{denis.grebenkov@polytechnique.edu}
\affiliation{Laboratoire de Physique de la Mati\`ere Condens\'ee, \\
CNRS -- Ecole Polytechnique, F-91128 Palaiseau, France }
\author{Dmitry Beliaev}
\affiliation{Mathematical Institute, University of Oxford, \\
Woodstock Road, Oxford OX2 6GG, UK}
\author{Peter W. Jones}
\affiliation{Yale University, Mathematics Department, \\
PO Box 208283, New Haven, CT 06520-8283, USA }



\begin{abstract}
We revise the L\'evy's construction of Brownian motion as a simple
though still rigorous approach to operate with various Gaussian
processes.  A Brownian path is explicitly constructed as a linear
combination of wavelet-based ``geometrical features'' at multiple
length scales with random weights.  Such a wavelet representation
gives a closed formula mapping of the unit interval onto the
functional space of Brownian paths.  This formula elucidates many
classical results about Brownian motion (e.g., non-differentiability
of its path), providing intuitive feeling for non-mathematicians.  The
illustrative character of the wavelet representation, along with the
simple structure of the underlying probability space, is different
from the usual presentation of most classical textbooks.  Similar
concepts are discussed for fractional Brownian motion,
Ornstein-Uhlenbeck process, Gaussian free field, and fractional
Gaussian fields.  Wavelet representations and dyadic decompositions
form the basis of many highly efficient numerical methods to simulate
Gaussian processes and fields, including Brownian motion and other
diffusive processes in confining domains.
\end{abstract}



\date{\today}


\maketitle

\section{Introduction}

Diffusion is a fundamental transport mechanism in nature and industry,
with applications ranging from physics to biology, chemistry,
engineering, and economics.  This process has attracted much attention
during the last decades, particularly in statistical and condensed
matter physics: diffusion-reaction processes; transport in porous
media and biological tissues; trapping in heterogeneous systems;
kinetic and aggregation phenomena like DLA, to name a few fields.
From an intuitive point of view, Brownian motion is often considered
as a continuous limit of lattice random walks.  However, a more
rigorous background is needed to answer subtle questions.  In
mathematical textbooks, Brownian motion is defined as an almost surely
continuous process with independent normally distributed increments
\cite{Revuz,Ito,Port,Bass,Borodin}.  The deceptive simplicity of this
definition relies on the notion of ``almost surely'' that, in turn,
requires a sophisticated formalism of Wiener measures in the space of
continuous functions, filtrations, sigma-algebra, etc.  Although this
branch of mathematics is well developed, it is rather difficult for
non-mathematicians, that is, the majority of scientists studying
Brownian motion in their every-day research.

In this paper, we discuss a different, but still rigorous, approach to
define and operate with Brownian motion as suggested by P. L\'evy
\cite{Levy}.  We construct from scratch a simple and intuitively
appealing representation of this process that gives a closed formula
mapping of the unit interval onto the functional space of Brownian
paths.  In this framework, sampling a Brownian path is nothing else
than picking up uniformly a point from the unit interval.
Figuratively speaking, Brownian motion is constructed here by adding
randomly wavelet-based geometrical features at multiple length scales.
The explicit formula elucidates many classical results about Brownian
motion (e.g., non-differentiability of its path).  The illustrative
character of the wavelet representation, along with the simple
structure of the underlying probability space, is different from the
usual presentation of most classical textbooks.

Among various amazing properties, Brownian motion is known to have a
self-similar structure: when a fragment of its path is magnified, it
``looks'' like the whole path.  In other words, any fragment obeys the
same probability law as the whole path.  As a consequence, Brownian
paths exhibit their features at (infinitely) broad range of length
scales.  As a matter of fact, multiscale geometrical structures are
ubiquitous in nature and material sciences \cite{Mandelbrot}.  For
instance, respiratory and cardiovascular systems start from large
conduits (trachea and artery) that are then split into thinner and
thinner channels, up to the size of few hundred microns for the
alveoli and several microns for the smallest capillaries
\cite{Weibel}.  Another example is a high-performance concrete which
is made with grains of different sizes (from centimeters to microns),
smaller grains filling empty spaces between larger ones.  The best
adaptive description of such self-similar structures relies on
intrinsicly multiscale functions to capture their mechanical or
transport properties at different length scales.  We illustrate this
idea by constructing Brownian motion using wavelets, a family of
functions with compact support and well defined scaling
\cite{Jaffard,Daubechies,Mallat}.  Wavelets appear as the natural
mathematical language to describe and analyze multiscale structures,
from heterogeneous rocks to biological tissues
\cite{Mehrabi97,Ebrahimi04,Friedrich11}.  The wavelet construction of
Brownian motion naturally extends to fractional Brownian motion and
other Gaussian processes and fields, allowing one to efficiently
simulate, for instance, turbulent diffusion with high Reynolds numbers
or financial markets.  In particular, we discuss the Gaussian (or
massless) free field and fractional Gaussian fields which appear as
basic models in different areas of physics, from astrophysics (cosmic
microwave background) to critical phenomena, quantum physics, and
turbulence \cite{Bardeen86,Kobayashi11,Fernandez,Dodelson}.  Written
in a spectral form in one dimension, Brownian motion and the
fractional Gaussian field look very similar, one of them being the
fractional derivative of the other.  Putting together these two
processes reveals deep relations between them, and this correspondence
carries over to higher dimensions.

Most importantly, the wavelet representation is a starting point for a
number of highly efficient numerical methods to simulate various
Gaussian processes and fields.  Though wavelet representations and the
related numerical methods are all known, they are not easily available
in a single source.  Indeed the totality of these methods seems to be
poorly understood, even amongst specialists.  The purpose of this
article is to present a unified and intuitive framework that is based
on elementary mathematical structures like, for example, dyadic
subdivision or wavelet tree.  We also present some simple
number-theoretic shortcuts and consequent numerical algorithms.
Finally, we discuss fast simulations of Brownian motion in confining
domains, where one of the difficulties is the ability to quickly
access the local geometry near the boundary.  These techniques can be
applied for studying Brownian motion and related processes or solving
partial differential equations in complex multiscale media.

We hope that this didactic article will provoke interesting
discussions amongst the experts and will help for a better
understanding of both theory and implementation of Brownian motion and
other Gaussian processes and fields for a much broader community of
their practical ``users'', namely, physicists, biologists, chemists,
engineers, and economists.

\section{Brownian motion}

In this section, we derive a wavelet representation of Brownian motion
in a simple explicit way, allowing one to gain an intuitive feeling of
this constructive approach.

\subsection{Physical view: upscaling and downscaling}
\label{sec:physical}

\begin{figure}
\begin{center}
\includegraphics[width=75mm]{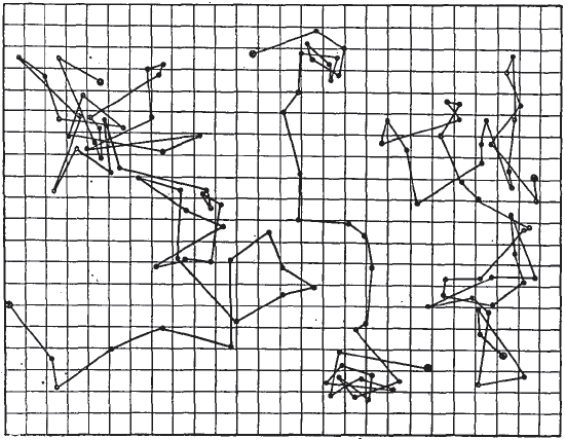}  
\end{center}
\caption{
Three random trajectories of small mastic grains in water recorded by
J. Perrin at 30-second intervals (reproduced from \cite{Perrin1909}).}
\label{fig:Perrin}
\end{figure}

In order to illustrate the basic idea of a wavelet representation, we
revisit the first single particle tracking experiment by R. Brown who
looked through a microscope at stochastic trajectories (now known as
Brownian paths) of pollens of Clarkia (primrose family)
\cite{Brown1828}.  First examples of such trajectories for mastic
grains in water were reported by J. Perrin
\cite{Perrin1908,Perrin1909}.  The essence of the wavelet
representation can be recognized in his description of three
trajectories (Fig. \ref{fig:Perrin}) that were recorded at 30-second
intervals \cite{Perrin1909}: ``{\it Ils ne donnent qu'une id\'ee
tr\`es affaiblie du prodigieux enchev\^etrement de la trajectoire
r\'eelle.  Si, en effet, on faisait des point\'es de seconde en
seconde, chacun de ces segments rectilignes se trouverait remplac\'e
par un contour polygonal de 30 c\^ot\'es relativement aussi
compliqu\'e que le dessin ici reproduit, et ainsi de suite.}''
\footnote{
``They provide only a very rough idea of the prodigious intricacy of
the real trajectory.  If one got points at every second, each of the
straight segments would be replaced by a polygonal contour of 30 sides
having approximately the same complexity as the current plot, and so
forth'' (translated by authors). }

Following this idea, let us record the positions of a particle (e.g.,
pollen or grain) at successive time moments with a selected time
resolution $\delta$.  Each particle submerged in water is permanently
``bombarded'' by water molecules.  Since the number of the surrounding
water molecules is very large and their tiny actions are mostly
uncorrelated, net microscopic displacements of the particle cannot be
considered deterministicly as in classical mechanics, but random.
Since the interaction of water molecules between them is very rapid as
compared to the macroscopic resolution scale, there is no memory
effect in their action on the heavy particle.  As a consequence, the
microscopic displacements of the particle are (almost) independent and
have a finite variance $\sigma^2$.  The macroscopic displacement
during the resolution time $\delta$ is the sum of a large number $N$
of these displacements with zero mean (no coherent flow).  Although
the average displacement is also zero, the stochastic fluctuations
around this value are of the order of $\sigma \sqrt{N}$.  Moreover,
the central limit theorem gives us a precise probabilistic description
of the fluctuations, resulting in a normal (or Gaussian) distribution
of macroscopic displacements of the particle \cite{Feller}.  It is
worth stressing that the Gaussian character of the macroscopic
displacements appears {\it without any specific knowledge about the
microscopic interactions}.  The only important information at
microscopic level was stationary, uncorrelated character of the
interaction, and finite variance (if one of these conditions is
missing, the resulting macroscopic process may exhibit anomalous
behavior, see \cite{Bouchaud90,Metzler00,Shlesinger99} and references
therein).  This is known as coarsening or upscaling: complex
interactions and the specific features of the underlying microscopic
dynamics are averaged out on macroscopic scales.  This is the reason
why Brownian motion (or diffusion in general) is so ubiquitous in
nature and science.  Once we know that the microscopic details are
irrelevant (under the conditions mentioned above), we can extend the
Gaussian behavior from macroscopic scales, where it has been
established, to microscopic scales.  This procedure can be called
downscaling, when we explicitly and purposefully transpose the
universal macroscopic behavior even onto smaller scales.  The
resulting {\it model} of the microscopic dynamics exhibits Gaussian
features at all scales.  While the true dynamics and its Gaussian
model can be completely different at microscopic scales, they become
identical at macroscopic scales.

Knowing that a particle moves continuously, we connect its successive
positions separated on time $\delta$ by a continuous line.  Since the
experimental setup is limited to the selected time scale $\delta$,
nothing can be said about the trajectory of the particle in between
two records.  In other words, the only condition for the trajectory to
pass through the recorded points leaves us a variety of choices for
the shape of the connecting continuous line.  The common choice is
connecting the successive positions by linear segments.  As one will
see, this choice fixes a particular wavelet representation, the Haar
wavelets.  We shall show that other wavelets, corresponding to other
choices of continuous connections, are as well useful.

When the magnification and time resolution of the experimental setup
are increased, smaller details of the particle's trajectory appear,
allowing one to refine the above piecewise linear approximation.
Repeating this procedure, in theory up to infinity, one recovers all
geometrical features and thus constructs the whole Brownian path.  In
what follows, we put this schematic description into a more rigorous
mathematical frame.

\subsection{Mathematical view: multiscale construction}

We start with the position ``records'' at every unit time: $t=1, 2, 3,
... $ (e.g., every second) along one coordinate (two- and higher
dimensional Brownian motion is then obtained by taking $d$ independent
copies of the one-dimensional process).  We focus on the time interval
between $0$ and $1$, the construction being applicable to any interval
$[\ell,\ell+1]$.  For convenience, Brownian motion is started at $t=0$
from the origin: $B(0)=0$.  By definition (or as a consequence of the
central limit theorem if one relies on the above physical reasoning),
the position at time $t=1$, $B(1)$, is a random variable $a_0$
distributed according to the standard normal (or Gaussian) law,
$\N(0,1)$, with mean zero and variance one
\begin{equation}
\label{eq:normal}
\P\{ a_0 \in (x,x+dx)\} = dx~ \frac{e^{-x^2/2}}{\sqrt{2\pi}}  .
\end{equation}
In physics, the variance $\sigma^2$ is related to the diffusion
coefficient, $D = \sigma^2/(2\tau)$, where $\tau$ is the one step
duration; here, $\sigma = \tau = 1$ yielding $D = 1/2$ throughout the
paper.  A linear approximation at this time scale ($\delta = 1$) is
simply
\begin{equation*}
B_0(t) = a_0 t ,
\end{equation*}
that connects the positions $B(0)=0$ and $B(1) = a_0$ by a linear
segment.

\begin{figure}
\begin{center}
\includegraphics[width=85mm]{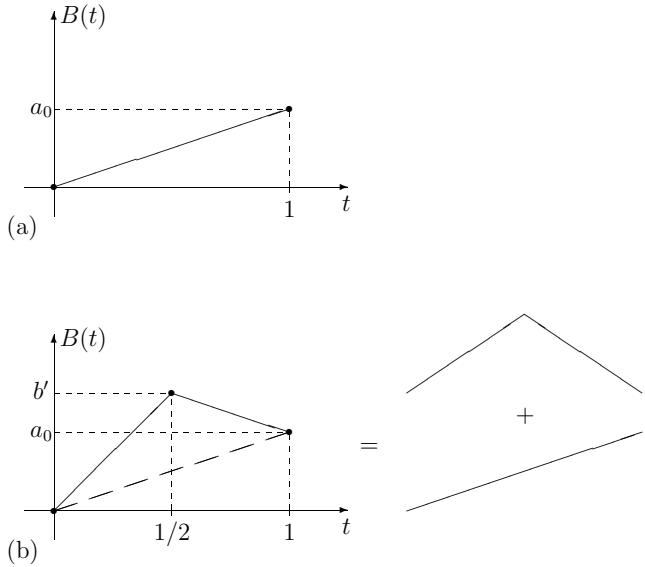}  
\end{center}
\caption{
Iterative construction of Brownian motion: {\bf (a)} linear
approximation $B_0(t)$ by a segment at scale $2^{0}$; {\bf (b)} linear
approximation $B_1(t)$ by two segments at scale $2^{-1}$.  The latter
``skewed tent'' can be uniquely represented as the sum of a
``symmetric tent'' and a linear shift.  }
\label{fig:tent}
\end{figure}

If the time resolution is doubled, a new, intermediate position $b' =
B(1/2)$ can now be seen (Fig. \ref{fig:tent}).  The random variable
$b'$ is conditioned by the fact that Brownian motion passes through
the points $(0,0)$ and $(1,a_0)$, the value of $a_0$ being already
known.  It is distributed according to the normal law with mean value
$\frac12(B(0)+B(1)) = a_0/2$ and variance $1/4$ (see
Appendix~\ref{sec:cond_law}).  In other words, one can write $b' =
a_0/2 + a_{00}/2$, where the new random variable $a_{00}\in \N(0,1)$
(i.e., distributed according to the standard normal law
(\ref{eq:normal})) is independent of $B(0) = 0$ and $B(1) = a_0$.

The linear approximation at the time scale $\delta = 1/2$ connects
three successive points $(0,0)$, $(1/2,b')$ and $(1,a_0)$ by two
linear segments:
\begin{equation}
B_1(t) = \begin{cases}  2t b',   \hskip 33mm  0\leq t\leq \frac12, \\
2t(a_0-b') + (2b' - a_0) ,  \quad \frac12 \leq t\leq 1 . \end{cases}
\end{equation}
The shape of this approximation looks like a skewed tent
(Fig. \ref{fig:tent}) that can be represented as a sum of a linear
shift and a ``symmetric tent'' function:
\begin{equation}
\label{eq:B1_decomp}
B_1(t) = a_0 t + (2b' - a_0) h_{00}(t) = a_0 t + a_{00} h_{00}(t) ,
\end{equation}
where the ``symmetric tent'' function $h_{00}(t)$ is 
\begin{equation}
h_{00}(t) = \begin{cases}  t ,  \hskip 13mm 0 \leq t \leq \frac12, \\
                1 - t,  \qquad \frac12 \leq t \leq 1 , \\
                0, \hskip 13mm  \textrm{otherwise}.  \end{cases}
\end{equation}
The decomposition (\ref{eq:B1_decomp}) into the linear function $t$
and the tent function $h_{00}(t)$ is unique.  The new approximation
$B_1(t)$ is obtained from the previous one, $B_0(t)$, by adding the
new term representing a smaller geometrical detail.

The same concept is applicable at every scale.  Assume that an
approximation $B_n(t)$ of Brownian motion is already constructed at
the time scale $\delta = 2^{-n}$, i.e., the positions $b_k = B(t_k)$
are known at successive times $t_k = k2^{-n}$, $k$ ranging from $0$ to
$2^n$.  The approximate Brownian path is a piecewise linear function
passing successively through these points.  

\begin{figure}
\begin{center}
\includegraphics[width=85mm]{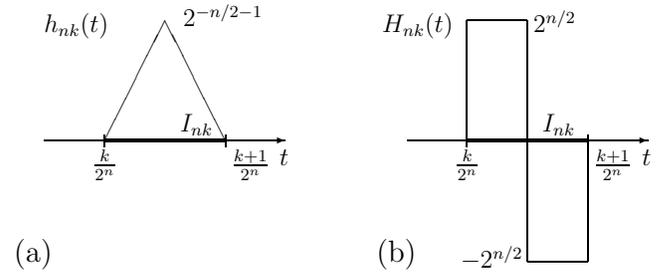}  
\end{center}
\caption{
Tent function $h_{nk}(t)$ {\bf (a)} and the related Haar function
$H_{nk}(t)$ {\bf (b)}, both having the support on the interval $I_{nk}
= [k2^{-n}, (k+1)2^{-n})$. }
\label{fig:haar}
\end{figure}

At the next time scale $2^{-n-1}$, a new, intermediate position $b' =
B(t')$ of Brownian motion at time $t' = (t_k+t_{k+1})/2 =
(k+1/2)2^{-n}$ should be determined for each $k$.  As previously, the
random variable $b'$ is conditioned by the fact that Brownian motion
is {\it known} to pass through the points $(t_k,b_k)$ and
$(t_{k+1},b_{k+1})$.  It is again distributed according to the normal
law, with mean value $(b_k + b_{k+1})/2$ and variance $2^{-n}/4$.  In
other words, one can write $b'$ as
\begin{equation}
\label{eq:b_prime}
b' = \frac12(b_k + b_{k+1}) + 2^{-n/2-1} a_{nk} ,
\end{equation}
where the new normal random variable $a_{nk} \in \N(0,1)$ is
independent of the other positions.  The linear segment between
$(t_k,b_k)$ and $(t_{k+1},b_{k+1})$ is then replaced by two linear
segments connecting the three successive points $(t_k,b_k)$,
$(t',b')$, and $(t_{k+1},b_{k+1})$.  This is a new ``skewed tent''
function which can be uniquely decomposed as the sum of the previous
linear segment (drift) and a symmetric tent function $h_{nk}(t)$ with
the weight $a_{nk}$, where
\begin{equation}
h_{nk}(t) = 2^{-n/2} h_{00}(2^n t - k)
\end{equation} 
is a rescaled symmetric tent function on the interval $I_{nk} =
[k2^{-n}, (k+1)2^{-n})$ (see Fig.~\ref{fig:haar}a).  We stress again
that the new approximation is obtained from the previous one by simply
adding the tent function $h_{nk}(t)$, representing a smaller
geometrical detail at the new scale $2^{-n-1}$, weighted by a normally
distributed coefficient $a_{nk}$ which is independent of the
previously determined positions.

This construction is applicable to all linear segments (all $k$) at
the given scale $n$, and it is valid for any scale.  Repeating this
procedure from the scale $2^0$ up to infinity, one obtains the Haar
wavelet representation of Brownian motion on the unit interval:
\begin{equation}
\label{eq:Bomega1}
B^\omega(t) = a_0^\omega t + \sum\limits_{n=0}^\infty \sum\limits_{k=0}^{2^n-1} a_{nk}^\omega h_{nk}(t) ,
\end{equation}
where all weights $a_0^\omega$ and $a_{nk}^\omega$ are independent
$\N(0,1)$ random variables.  Here we introduced the superscript
$\omega$ in order to stress that a sampling of Gaussian weights
$a_{nk}$ yields a random realization of Brownian motion.  We will
discuss in Sect. \ref{sec:weights} that all these Gaussian weights can
be constructed from a single random number $\omega$ from the unit
interval that provides a natural parameterization of Brownian paths.

The dyadic structure of the intervals implies that for any $n\in \Z$,
there exists only one interval $I_{nk}$ of length $2^{-n}$ containing
the point $t$: $k2^{-n}\leq t<(k+1)2^{-n}$.  The index $k$ is simply
the integer part of $2^n t$: $k = \lfloor 2^n t \rfloor$ (i.e., the
largest integer that does not exceed $2^n t$).  As a consequence, the
convergence in the above formula is very rapid.  In fact, if one needs
to obtain the value of function $B^\omega(t)$ with a desired precision
$\ve$, it is sufficient to calculate the first $\log_2(1/\ve)$ terms,
$\log_2 x$ being the logarithm of $x$ on the base of $2$.  

Subtracting the linear term $a_0^\omega t$ from Eq. (\ref{eq:Bomega1})
yields the Haar wavelet representation of a Brownian bridge on the
unit interval, i.e., Brownian motion conditioned to return to $0$ at
time $t=1$.

\subsection{Dyadic decomposition and interval subdivision}

\begin{figure}
\begin{center}
\includegraphics[width=75mm]{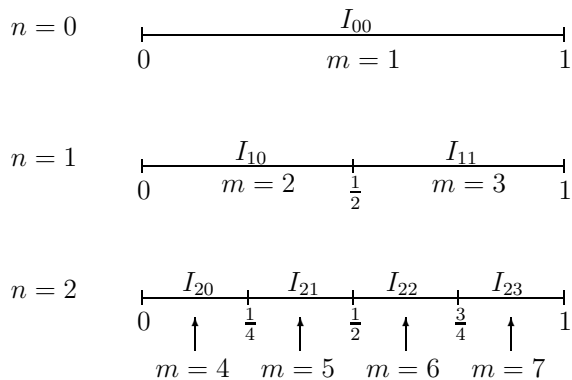}  
\end{center}
\caption{
Enumeration of the dyadic subintervals by a single index $m$. }
\label{fig:intervals}
\end{figure}

In the wavelet representation (\ref{eq:Bomega1}), the first sum is
carried over all scales $n$, while the second sum covers all $2^n$
subintervals $I_{nk}$ at the scale $n$.  In some cases, it is
convenient to enumerate all the dyadic subintervals $I_{nk}$ by a
single index $m = 2^n + k$ as shown on Fig.~\ref{fig:intervals}.
Since $k$ is ranging from $0$ to $2^{n-1}$, the new index $m$ uniquely
identifies the interval $I_{nk}$.  In particular, one easily retrieves
the pair $(n,k)$ from $m$ as
\begin{equation*}
n = \lfloor \log_2 m \rfloor ,   \hskip 10mm  k = m - 2^n .
\end{equation*}
Using the notations
\begin{equation*}
\tilde{a}_m^\omega = \begin{cases} ~ a_0^\omega , \hskip 3mm  m = 0 , \\
                         a_{nk}^\omega , ~~  m > 0 , \\
\end{cases}  \hskip 5mm
\tilde{h}_m(t) = \begin{cases}  ~~~t ,  \hskip 7mm  m = 0 , \\
                     h_{nk}(t) , ~~ m > 0 ,\\
\end{cases}
\end{equation*}
we can write Eq.~(\ref{eq:Bomega1}) in a more compact form
\begin{equation}
\label{eq:Bomega2}
B^\omega(t) = \sum\limits_{m=0}^\infty \tilde{a}_m^\omega ~ \tilde{h}_m(t) .
\end{equation}
As a result, Brownian trajectory is decomposed into a sum of tent
functions (plus a linear drift) with random independent identically
distributed Gaussian weights.  As illustrated below, all these weights
can be determined from a single uniformly distributed random variable.

\subsection{Representation of Gaussian weights}
\label{sec:weights}

In this subsection, we illustrate how all random Gaussian weights
$\tilde{a}^\omega_m$ can be explicitly related to a single uniformly
distributed random number $\omega$.  In other words, we show that the
complicated abstract probability space of Brownian paths can have a
simple parameterization.  However, this construction is not relevant
from practical point of view, and the subsection can be skipped at
first reading.

\begin{figure}
\begin{center}
\includegraphics[width=75mm]{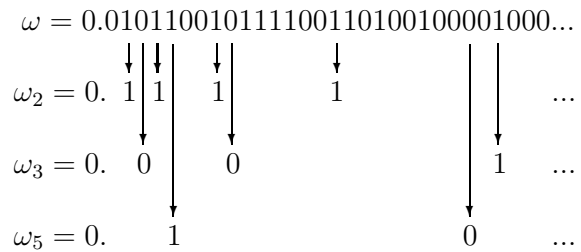}  
\end{center}
\caption{
Generation of an infinite sequence of independent uniformly
distributed variables $\omega_2$, $\omega_3$, $\omega_5$, $\ldots$
using binary expansion of a single number $\omega$ from the unit
interval.  For instance, $\omega_2$ is constituted of the $2^{\rm
nd}$, $4^{\rm th}$, $8^{\rm th}$, $\ldots$ bits of $\omega$. }
\label{fig:omega}
\end{figure}

We consider the binary expansion of a given real number $\omega$ from
the unit interval
\begin{equation}
\omega = 0.b_1b_2b_3b_4\ldots
\end{equation}
where $b_i$ are equal to $0$ or $1$ (note that $\omega = 1$ is
expanded as $0.111\ldots$ instead of its equivalent form
$1.000\ldots$).  A uniform picking up of $\omega$ in $[0,1]$ is
equivalent to {\it independent} random choice of its binary digits (or
bits) $b_i$.  Then we choose a prime number $p$ and construct another
number $\omega_p$ using the bits of $\omega$ at positions $p$, $p^2$,
$p^3$, $\ldots$
\begin{equation}
\omega_p = 0.b_p b_{p^2} b_{p^3} b_{p^4}\ldots
\end{equation}
For example, $\omega_2 = 0.b_2 b_4 b_8 b_{16}\ldots$, $\omega_3 =
0.b_3b_9b_{27}b_{81}\ldots$, etc. (Fig.~\ref{fig:omega}).  If $p$ and
$q$ are two different prime numbers then $\omega_p$ and $\omega_q$ are
independent as being constructed from separate sets of independent
bits $b_i$.  Moreover, if $\omega$ is chosen uniformly from the unit
interval, then each $\omega_p$ is also uniformly distributed on the
unit interval.  Consequently, a single random number $\omega$ gives
rise to an infinite sequence $\{\omega_p\}$ of independent uniformly
distributed random variables.  In a more formal way, the real numbers
$\omega_p$ can be written as
\begin{equation}
\label{eq:omegap}
\omega_p = \sum\limits_{n=1}^\infty 2^{-n-1}\bigl(1 + R(2^{p^n}\omega)\bigr) ,
\end{equation}
where $R(x) = (-1)^{\lfloor x \rfloor}$ is the Rademacher function.

At last, we need to pass from uniformly distributed to normally
distributed variables.  For this purpose, we define the inverse
$\Phi(x)$ of the error function: for $x\in[0,1]$, the value $y$ of the
function $\Phi(x)$ satisfies
\begin{equation}
\label{eq:erf_inv}
\frac{1}{\sqrt{2\pi}}\int\limits_{-\infty}^y dz ~e^{-z^2/2} = x .
\end{equation}
Although there is no simple analytic form for the function $\Phi(x)$,
many properties can be easily derived, and the whole function can be
tabulated with any required precision.

If $p_m$ denotes the $(m+1)^{\rm th}$ prime (e.g., $p_0=2$, $p_1=3$,
$p_2=5$), then we set
\begin{equation}
\tilde{a}_m^\omega = \Phi(\omega_{p_m}),   \hskip 8mm  m = 0,1,2,...
\end{equation}
By construction, $\tilde{a}_m^\omega$ are independent normally
distributed random variables.  In other words, Eqs. (\ref{eq:omegap},
\ref{eq:erf_inv}) map a uniformly distributed $\omega$ onto a sequence
of Gaussian weights $\tilde{a}_m$.  As a result, $B^\omega(t)$ is
constructed as a mapping from the unit interval, $\omega\in [0,1]$,
onto the space of real-valued functions (more precisely, the H\"older
space $H_{1/2-\epsilon}$ with any $\ve > 0$).  In other words, any
Brownian path is explicitly encoded by the real number $\omega$.
Picking up the real number $\omega$ from the unit interval (with
uniform measure) is thus equivalent to choosing a Brownian path (with
Wiener measure).  In this representation, the probability space for
Brownian motion is nothing else than the unit interval with uniform
measure.  It is intuitively much simpler than the classical
construction of the probability space by means of Wiener measure,
filtrations, etc.  At the same time, this mapping is evidently neither
continuous, nor injective (e.g., two numbers $\omega$ and $\omega'$
that differ at 6-th bit correspond to the same sequence of Gaussian
random numbers).  The mapping remains a rather formal construction
whose only purpose was to show the equivalence between two spaces.

\subsection{Haar wavelets}

\begin{figure}
\begin{center}
\includegraphics[width=80mm]{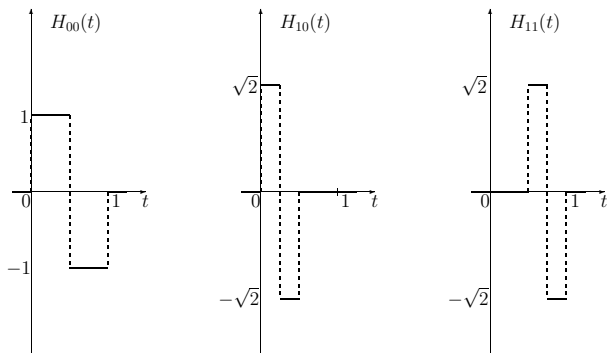} 
\end{center}
\caption{
Haar wavelets $H_{nk}(t)$ are obtained by dilations and translations
of the mother function $\phi^{1,1}(t) = H_{00}(t)$ (shown on the
left). }
\label{fig:haars}
\end{figure}

In Eqs.~(\ref{eq:Bomega1}) or (\ref{eq:Bomega2}), Brownian motion is
decomposed into a sum of a linear function and tent functions.
Figure~\ref{fig:haar}a illustrates that any tent function $h_{nk}(t)$
can actually be represented as the integral of a piecewise-constant
function
\begin{equation}
h_{nk}(t) = \int\limits_0^t dt' ~H_{nk}(t') ,
\end{equation}
where $H_{nk}(t)$ is called the Haar function and defined to be $0$ on
the complement of $I_{nk}$ and to take values $2^{n/2}$ and $-2^{n/2}$
on its left and right subintervals, respectively
(Fig.~\ref{fig:haar}b).  In fact, all Haar functions are obtained by
translations and dilations of a single ``mother'' function
$\phi^{1,1}(t)$:
\begin{equation}
\label{eq:phi11}
\phi^{1,1}(t) = \begin{cases} ~~1, ~~~~ 0 \leq t < \frac12 , \\
-1, \quad \frac12 < t \leq 1 ,\\
~~0 ,  ~~~~ \textrm{otherwise}. \\
\end{cases}
\end{equation}
As illustrated on Fig.~\ref{fig:haars}, one has
\begin{equation}
\label{eq:Haar_phi}
H_{nk}(t) = 2^{n/2} \phi^{1,1}(2^n t - k) .
\end{equation}

It is convenient to use previously introduced single index $m$ to
denote different Haar functions:
\begin{equation*}
\tilde{H}_m(t) = \begin{cases} ~~ 1, \hskip 8mm  m = 0, \\
                  H_{nk}(t), ~~ m > 0. \\
\end{cases}
\end{equation*}
Eq.~(\ref{eq:Bomega2}) yields the following representation for
Brownian motion
\begin{equation}
\label{eq:Bnoise}
B^\omega(t) = \int\limits_0^t dW^\omega(t') ,
\end{equation}
where $dW^\omega(t)$ denotes the Gaussian white noise which is defined
here through the Haar wavelet decomposition
\begin{equation}
\label{eq:Bnoise2}
dW^\omega(t) = \sum\limits_{m=0}^\infty \tilde{a}_m^\omega ~\tilde{H}_m(t) .
\end{equation}

It is easy to check that Haar functions $\{\tilde{H}_m(t)\}$ (together
with a constant function) form an orthonormal basis in the space
$L^2([0,1])$ of measurable and square integrable functions that is
\begin{equation*}
\int\limits_0^1 dt ~ \tilde{H}_m(t) \tilde{H}_{m'}(t) = \delta_{m,m'} .
\end{equation*}
Moreover, this basis is known to be complete in $L^2([0,1])$.  This
means that any function from $L^2([0,1])$ can be decomposed into a
linear combination of the Haar functions (and a constant).  From now
on, we drop the superscript $\omega$ for getting simpler notations
although the parameterization by $\omega$ remains valid for all
discussed processes.

\subsection{General spectral representation}

The wavelet representation (\ref{eq:Bnoise2}) can be extended to {\it
any} complete orthonormal basis $\{\psi_i(t)\}$ of $L^2([0,1])$.  In
fact, if the orthonormal basis $\{\psi_i(t)\}$ is complete, one can
decompose any function $\tilde{H}_m(t)$ into a linear combination of
$\psi_i(t)$:
\begin{equation*}
\tilde{H}_m(t) = \sum\limits_{i=0}^\infty c_{m,i}~ \psi_i(t) ,
\end{equation*}
where the coefficients $c_{m,i}$ satisfy the orthogonality relation
\begin{equation}
\label{eq:cmi}
\sum\limits_{m=0}^\infty c_{m,i}^2 = 1 \qquad (i = 0,1,2,\ldots).
\end{equation}
Substitution of this decomposition into Eq.~(\ref{eq:Bnoise2}) gives
\begin{equation}
\label{eq:dW_psi}
dW(t) = \sum\limits_{m=0}^\infty \tilde{a}_m \sum\limits_{i=0}^\infty c_{m,i}~ \psi_i(t) = 
\sum\limits_{i=0}^\infty \hat{a}_i ~ \psi_i(t) ,
\end{equation}
with new random weights
\begin{equation*}
\hat{a}_i = \sum\limits_{m=0}^\infty \tilde{a}_m ~c_{m,i} .
\end{equation*}
The sum of independent Gaussian variables is a Gaussian variable, and
its variance is simply the sum of the squared coefficients
$c_{m,i}^2$, which is equal to $1$ according to Eq.~(\ref{eq:cmi}).
In other words, $\hat{a}_i \in \N(0,1)$.  Moreover, the new random
variables $\hat{a}_i$ are independent due to the orthogonality of the
functions $\psi_i(t)$.  We have thus shown that the Gaussian white
noise can be decomposed into a linear combination with independent
Gaussian weights in {\it arbitrary} complete orthonormal basis of
$L^2([0,1])$.

The completeness of the basis $\{\psi_i(t)\}$ yields the usual
covariance of the Gaussian white noise
\begin{equation*}
\begin{split}
\E\{ dW(t_1) dW(t_2)\} & = \sum\limits_{i_1,i_2=0}^\infty \psi_{i_1}(t_1) \psi_{i_2}(t_2) \E\{\hat{a}_{i_1} \hat{a}_{i_2}\} \\
& =  \sum\limits_{i=0}^\infty \psi_{i}(t_1) \psi_{i}(t_2) = \delta(t_1-t_2), \\
\end{split}
\end{equation*}
where $\delta$ is the Dirac distribution (or ``$\delta$-function'').

Substituting Eq. (\ref{eq:dW_psi}) into Eq. (\ref{eq:Bnoise}), one
gets
\begin{equation}
\label{eq:Bomega4}
B(t) =  \sum\limits_{i=0}^\infty \hat{a}_i \int\limits_0^t dt' ~\psi_i(t') ,
\end{equation}
For instance, one can consider the Fourier basis on the unit interval,
\begin{equation*}
\psi_i(t) = \sqrt{2}~ \cos (\pi(i-1/2)t) ,
\end{equation*}
in order to get the Karhunen-Lo\`eve expansion of Brownian motion
\cite{Loeve}:
\begin{equation}
\label{eq:B_Fourier}
B(t) = \sqrt{2}\sum\limits_{i=0}^\infty \hat{a}_i ~\frac{\sin(\pi (i-1/2)t)}{(i-1/2)\pi} .
\end{equation}

\subsection{Basic properties of Brownian motion}

We have explicitly constructed the Haar wavelet decomposition
(\ref{eq:Bomega1}) and then general representation (\ref{eq:Bomega4})
in order to reproduce the basic properties of Brownian motion.
Alternatively, one could first postulate such a representation as a
definition of Brownian motion and then check that the basic properties
are fulfilled.  To illustrate this point, we check several properties.

(i) Brownian motion is a Gaussian process with independent increments.
First, $B(t)$ is Gaussian as being a linear combination
(\ref{eq:Bomega4}) of Gaussian variables.  Let $t_1 < t_2$ and $t_3 <
t_4$ define two increments, $B(t_2) - B(t_1)$ and $B(t_4) - B(t_3)$.
If $t_2 \leq t_3$ (i.e., the increments do not ``overlay''), then they
are independent (similar statement holds if $t_4 \leq t_1$ by
symmetry).  To proof this statement, we decompose the unit interval as
\begin{equation}
[0,1] = [0,t_1)\cup [t_1,t_2] \cup (t_2,t_3) \cup [t_3,t_4] \cup (t_4,1],
\end{equation}
(if one of subintervals $[0,t_1)$, $(t_2,t_3)$ or $(t_4,1]$ is empty,
it can be ignored).  The basis $\{\psi_i(t)\}$ in
Eq. (\ref{eq:Bomega4}) can be chosen as the direct product of the Haar
eigenbases on each subinterval.  For instance,
$\{\psi_i^{[t_1,t_2]}\}$ is the Haar basis of $L^2([t_1,t_2])$ which
is extended to $[0,1]\backslash [t_1,t_2]$ by zeros.  In this
particular representation, one has
\begin{equation}
\label{eq:B_increment_aux1}
\begin{split}
& B(t_2) - B(t_1) = \int\limits_{t_1}^{t_2} dt' \sum\limits_{i=0}^\infty \hat{a}_i ~ \psi_i(t') \\
& = \int\limits_{t_1}^{t_2} dt' \sum\limits_{i=0}^\infty \hat{a}_i^{[t_1,t_2]} ~ \psi_i^{[t_1,t_2]}(t') = \hat{a}_0^{[t_1,t_2]}~ \sqrt{t_2 - t_1} .\\
\end{split}
\end{equation} 
In the second equality, the Haar functions from other subintervals
(except $[t_1,t_2]$) vanished by construction.  In turn, all the Haar
functions $\psi_i^{[t_1,t_2]}(t')$ on $[t_1,t_2]$ (with $i>0$)
vanished after integration due to their orthogonality to the constant.
The only remaining contribution is the constant term which has the
unit $L^2([t_1,t_2])$ norm: $\psi_0^{[t_1,t_2]}(t) = (t_2 -
t_1)^{-1/2}$.  Integrating this term, one gets the right-hand side of
Eq. (\ref{eq:B_increment_aux1}) which shows that the increment $B(t_2)
- B(t_1)$ is a Gaussian variable with mean zero and variance
$t_2-t_1$, as expected.  Moreover, the same representation for
$[t_3,t_4]$ yields
\begin{equation}
\label{eq:B_increment_aux2}
\begin{split}
& B(t_4) - B(t_3) = \int\limits_{t_3}^{t_4} dt' \sum\limits_{i=0}^\infty \hat{a}_i ~ \psi_i(t') \\
& =  \int\limits_{t_3}^{t_4} dt' \sum\limits_{i=0}^\infty \hat{a}_i^{[t_3,t_4]} ~ \psi_i^{[t_3,t_4]}(t') = \hat{a}_0^{[t_3,t_4]}~ \sqrt{t_3 - t_4} , \\
\end{split}
\end{equation} 
and the random weights $\hat{a}_0^{[t_1,t_2]}$ and
$\hat{a}_0^{[t_3,t_4]}$ are independent by construction.  As a
consequence, the increments $B(t_2) - B(t_1)$ and $B(t_4) - B(t_3)$
are independent.

(ii) The mean and covariance of Brownian motion are:
\begin{equation}
\E\{B(t)\} = 0,  \qquad  \E\{B(t_1)B(t_2)\} = \min\{t_1,t_2\}.  
\end{equation}
The first statement is obvious from Eq. (\ref{eq:Bomega4}) given that
all weights have mean zero.  To prove the second statement, one
assumes that $t_2 > t_1$ and considers
\begin{equation*}
\begin{split}
\E\{B(t_1)B(t_2)\} & = \E\{(B(t_1) - B(0))(B(t_2) - B(t_1)\} \\
& + \E\{ [B(t_1)-B(0)]^2\} .  \\
\end{split}
\end{equation*}
The first term vanishes due to the independence of increments (and
$B(0) = 0$), while the second term is equal to $t_1$ according to
Eq. (\ref{eq:B_increment_aux1}).

(iii) Brownian motion is continuous but nowhere differentiable almost
surely.  The proof relies on a simple fact that the Gaussian weights
$a_{nk}$ cannot be too large, e.g., the probability that $|a_{nk}| >
n$ decays extremely fast (as $e^{-n^2/2}$ for large enough $n$).  In
turn, the norm of tent functions decreases exponentially that ensures
the continuity of Brownian motion and the convergence of a partial sum
approximation in Eq. (\ref{eq:Bomega1}) or similar expressions to
Brownian motion.  Moreover, the remainder of this approximation
decreases exponentially fast with the truncated scale $N$ (for
technical details, see Appendix \ref{sec:continuity}).

\subsection{Alpert-Rokhlin wavelets}

As we mentioned in Sect.~\ref{sec:physical}, taking the particular
orthonormal basis is equivalent to choosing a way to connect
successive positions of Brownian motion at a finite scale $\delta$.
The Haar wavelets and the resulting tent functions present the
simplest way of connection by linear segments.  Such a piecewise
linear approximation of Brownian motion introduces singularities at
the connection nodes (corners).  For some problems, it is convenient
to deal with a smooth approximation of Brownian motion at finite
scales (although a true Brownian trajectory, the limiting curve,
remains nowhere differentiable).  For this purpose, one can use the
Alpert-Rokhlin multiwavelet basis
\cite{Alpert,Alpert92,Beylkin91,Beylkin92}.  This basis is generated
by a set of $q$ functions $\phi^{q,1}(t), \ldots, \phi^{q,q}(t)$ which
are supported on the interval $[0,1]$, are piecewise polynomials of
degree $q-1$ on $[0,1/2]$ and on $[1/2,1]$, and satisfy the moment
cancellation conditions
\begin{equation}
\label{eq:moments}
\int\limits_0^1 dt~ t^k~ \phi^{q,p}(t) = 0, ~~~ \begin{array}{l} k = 0, 1, \ldots, q-1 , \\ p = 1, 2, \ldots, q .\\ \end{array} 
\end{equation}
These mother functions generate the Alpert-Rokhkin multiwavelets of
order $q$ by translations and dilations:
\begin{equation*}
\phi^{q,p}_{nk}(t) = 2^{-n/2} \phi^{q,p}(2^n t - k)  \qquad 
\begin{array}{l} n = 0,1,2,\ldots, \\ k = 0,1,2,\ldots 2^n-1. \\ \end{array}
\end{equation*}
The set of functions $\{\phi^{q,p}_{nk}(t)\}$, completed by the set of
orthonormal polynomials of order $m < q$, forms a complete basis of
$L^2([0,1])$.  This completion is necessary because all mother
functions $\phi^{q,p}(t)$ (and thus all $\phi^{q,p}_{nk}(t)$) are
orthogonal by construction to all polynomials of order $m < q$.
Similarly, the Haar wavelets were completed by a constant function.

\begin{figure}
\begin{center}
\includegraphics[width=42mm]{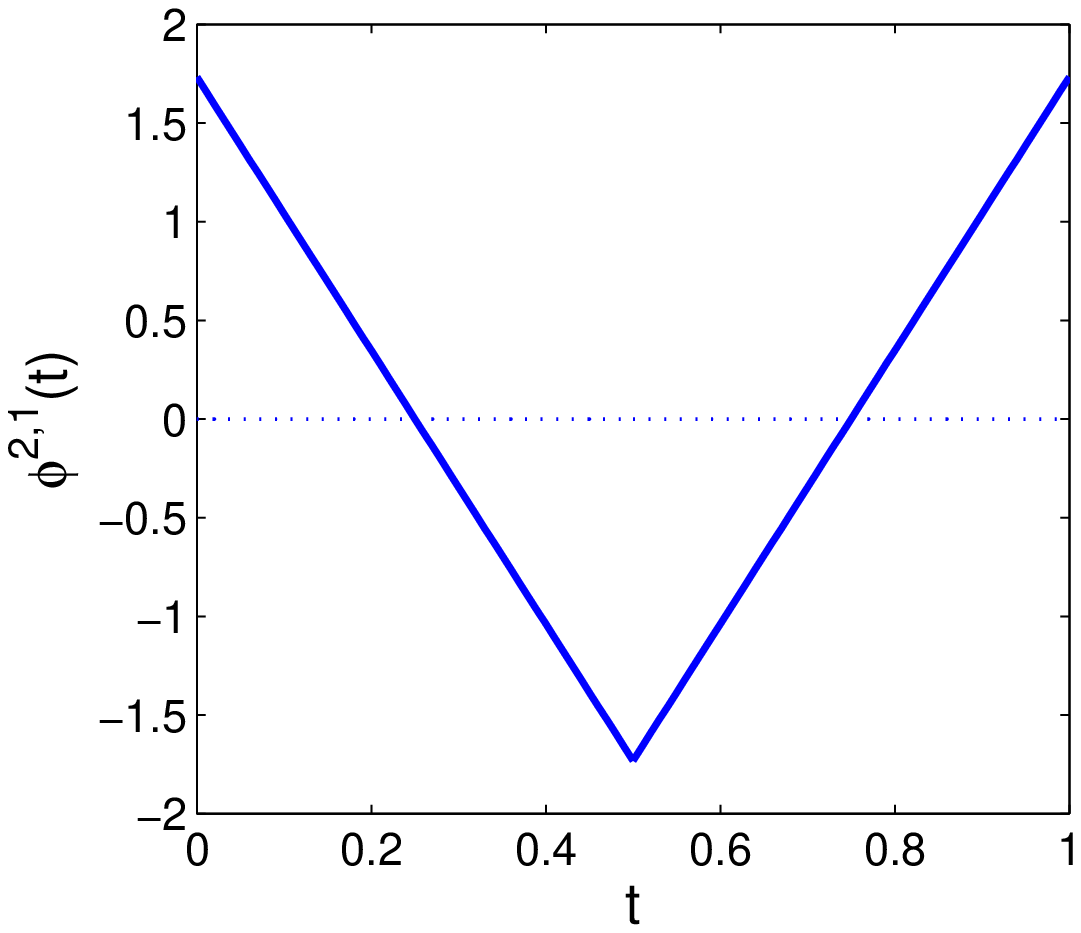} 
\includegraphics[width=42mm]{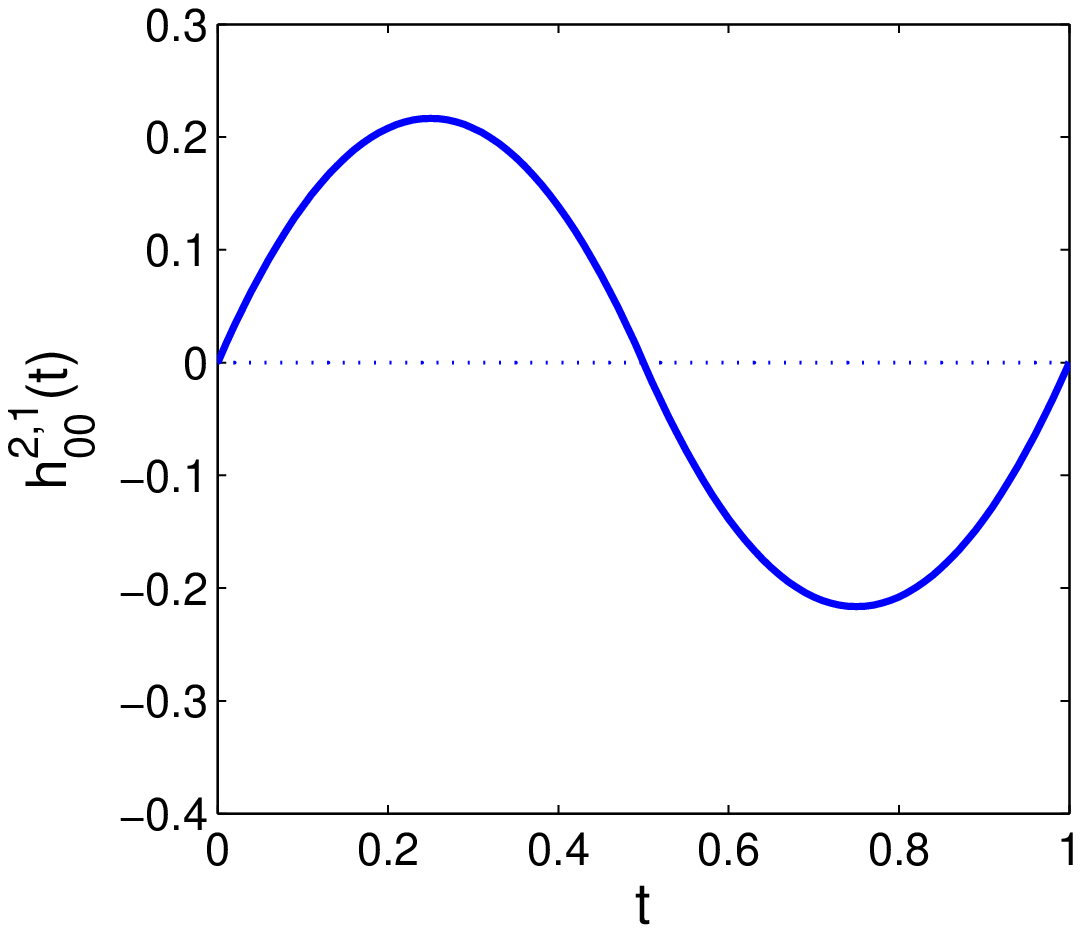} 
\includegraphics[width=42mm]{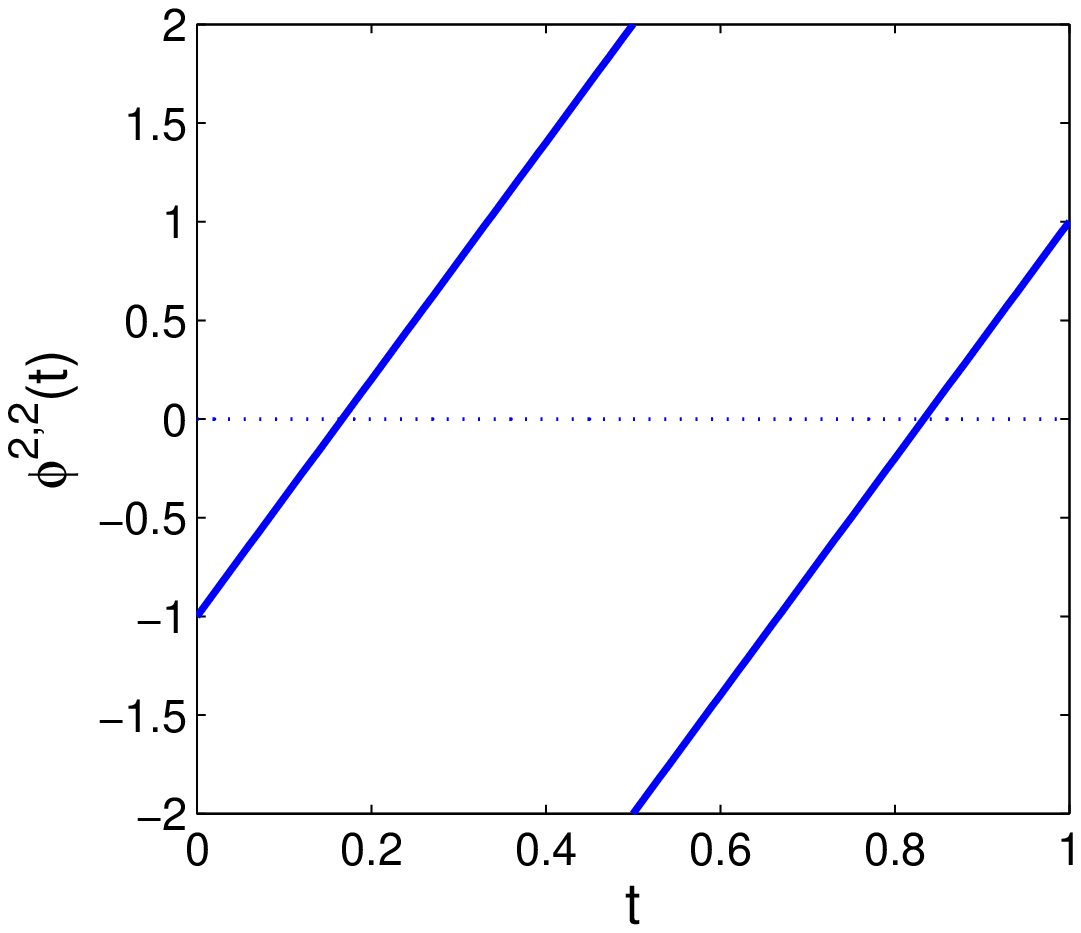} 
\includegraphics[width=42mm]{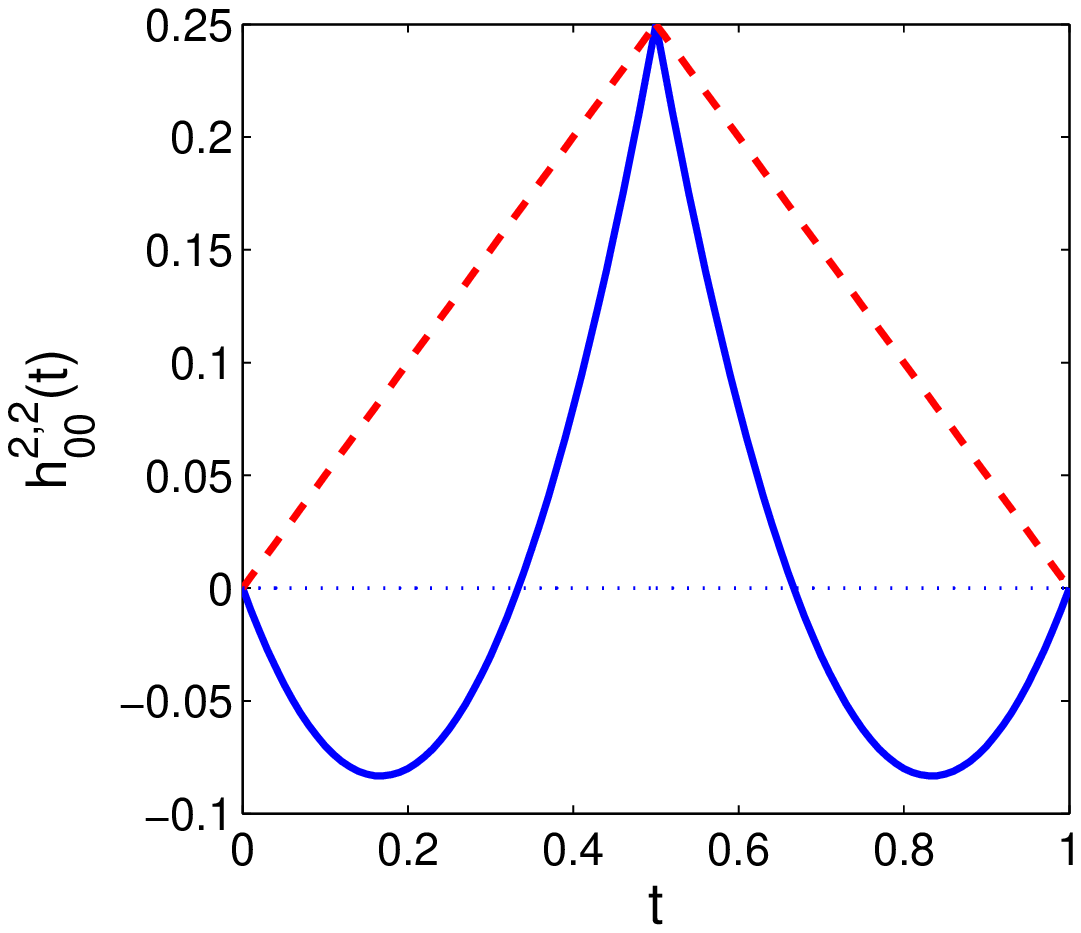} 
\end{center}
\caption{
Two mother functions for Alpert-Rokhlin first-order multiwavelets (on
the left) and their integrals $h^{q,p}_{00}(t)$ (on the right).  The
use of these wavelets corresponds to another type of connection (not
by linear segments) between successive points of Brownian motion in
the refinement procedure.  For comparison, the tent function
$h_{00}(t)$ is shown on the last plot by dashed line.  Note that
horizontal and vertical scales are not matched here. }
\label{fig:alpert}
\end{figure}

When $q = 1$, there is only one mother function $\phi^{1,1}(t)$
defined by Eq.~(\ref{eq:phi11}) which generates the Haar wavelets by
translations and dilations.  For $q = 2$, there are two mother
functions (Fig.~\ref{fig:alpert}), satisfying the moment cancellation
conditions (\ref{eq:moments}):
\begin{equation*}
\begin{split}
\phi^{2,1}(t) & = \begin{cases} \sqrt{3}~(1 - 4t),  \qquad   0 \leq t < \frac12, \\
\sqrt{3}~ (4t - 3),  \qquad \frac12 < t \leq 1, \\
~~~~~~~ 0 ,  \hskip 15mm \textrm{otherwise}  \\  \end{cases}  \\
\phi^{2,2}(t) & = \begin{cases} 6t - 1 , \qquad  0 \leq t < \frac12, \\
6t - 5 , \qquad  \frac12 < t \leq 1, \\
~~~~ 0  , \hskip 10mm \textrm{otherwise}  \\  \end{cases} \\
\end{split}
\end{equation*}
Higher-order mother functions (with $q > 2$) can be constructed
through an orthogonalization procedure (see \cite{Alpert,Elliott94}
for details and examples).

Note that the wavelet representation of Brownian motion involves the
integral of wavelets
\begin{equation}
h^{q,p}_{00}(t) = \int\limits_0^t dt' \phi^{q,p}(t') .
\end{equation}
For instance, one gets for $q = 2$ (Fig. \ref{fig:alpert})
\begin{eqnarray*}
h^{2,1}_{00}(t) & = & \begin{cases}  \sqrt{3}~ t~ (1 - 2t), \hskip 12mm 0 \leq t < \frac12, \\
\sqrt{3}~ (1 - t)(1 - 2t) , \quad \frac12 < t \leq 1, \\  \end{cases} \\
h^{2,2}_{00}(t) & = & \begin{cases}  t ~(3t - 1), \hskip 12mm   0 \leq t < \frac12, \\
(t - 1)(3t - 2) , \quad \frac12 < t \leq 1. \\  \end{cases}
\end{eqnarray*}
while the other functions $h^{2,1}_{nk}(t)$ and $h^{2,2}_{nk}(t)$ are
obtained by dilations and translations.  As a consequence, Brownian
motion gets a closed formula in terms of the Alpert-Rohklin
multiwavelets of order $2$:
\begin{equation}
B(t) = a_0 t + a_1 \sqrt{3} t(t-1) + \sum\limits_{n=0}^\infty \sum\limits_{k=0}^{2^n-1} \sum\limits_{p=1}^2 a_{nk}^{(p)} h^{2,p}_{nk}(t) ,
\end{equation}
where all weights $a_0$, $a_1$, $a_{nk}^{(p)}$ are independent
$\N(0,1)$ variables, and the second term is the integral of the linear
basis function $\sqrt{3}(2t-1)$.  An extension of this representation
to the Alpert-Rohklin multiwavelet basis of order $q$ is
straightforward.

\subsection{Numerical implementation}

\begin{figure}
\begin{center}
\includegraphics[width=85mm]{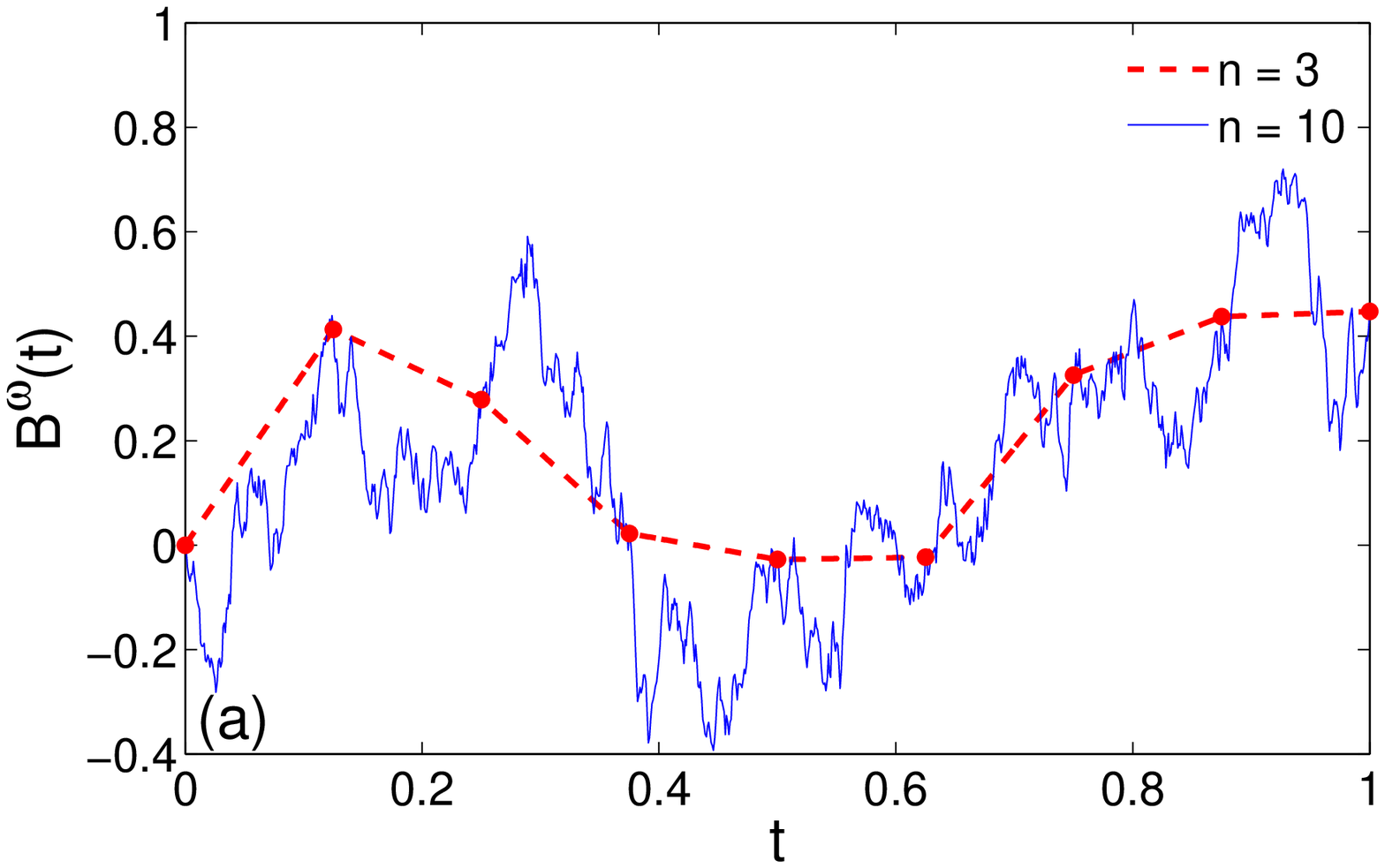} 
\includegraphics[width=85mm]{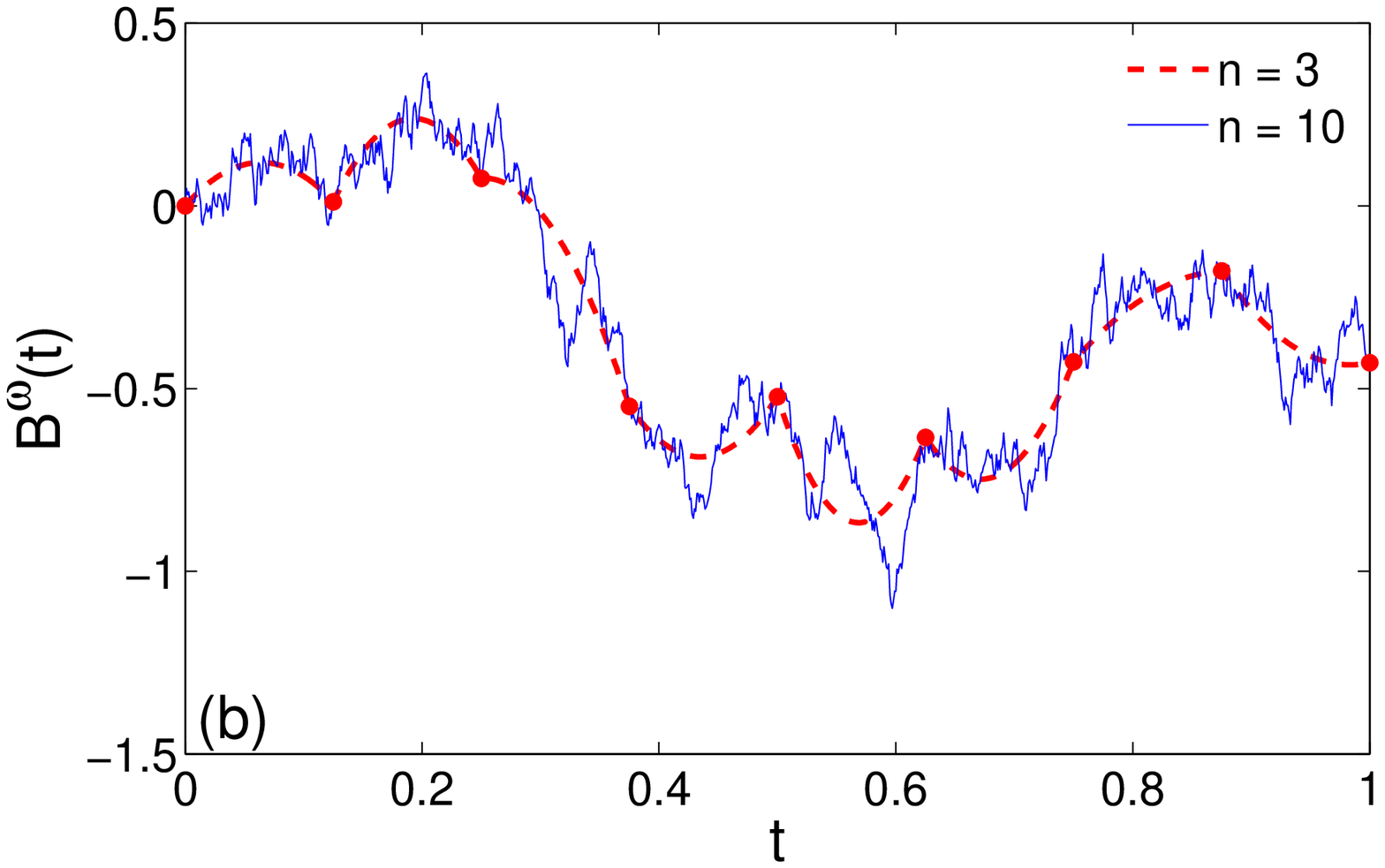} 
\end{center}
\caption{
A random Brownian path at scales $n=3$ (dashed line), and $n = 10$
(solid line) with Haar wavelets {\bf (a)} and Alpert-Rohkin wavelets
with $q = 2$ {\bf (b)}. }
\label{fig:BM_iterates}
\end{figure}

The wavelet representation of Brownian motion can be easily
implemented in practice.  To carry computations with a (fixed) desired
precision $\ve$, it is sufficient to truncate the first sum in
Eq.~(\ref{eq:Bomega1}) or equivalent relation up to $N =
\lfloor \log_2(1/\ve) \rfloor$, because higher-order terms describe geometrical
details at smaller scales.  The remainder of this series can be
estimated using Eq. (\ref{eq:approx_remainder}).  For Haar wavelets,
such a truncation qualitatively corresponds to an approximation of
Brownian motion by a broken line composed of linear segments of length
close to $\ve$, while the Alpert-Rokhlin wavelets yield smoother
approximations (Fig. \ref{fig:BM_iterates}).  This is a fascinating
feature of wavelets, allowing one to capture geometrical details at
different scales.

A realization of a Brownian path is completely determined by a set of
random coefficients $a_{nk}^\omega$ (or $\tilde{a}_m^\omega$).  In
Sect.~\ref{sec:weights}, we discussed an explicit scheme to generate
all these coefficients from a randomly chosen number $\omega$ from the
unit interval.  In practice, one can use standard routines to generate
pseudo-random normally distributed weights $a_{nk}^\omega$ (or
$\tilde{a}_m^\omega$).  The computation of tent functions $h_{nk}(t)$
can be easily implemented.  Consequently, the computation of $B(t)$ at
any time $t$ requires only $\log_2(1/\ve)$ operations, each of them
consisting of finding $h_{nk}(t)$, multiplying it by $a_{nk}^\omega$,
and summing their contributions.

It is instructive to compare the wavelet approach to conventional
techniques.  We consider the computation of all positions $b_k =
B(t_k)$ at equidistant times $t_k = k2^{-N}$ ($k=0,...,2^N$) at some
scale $\ve = 2^{-N}$.  In a classical scheme, Brownian motion is
modeled by a sequence of small random jumps
\begin{equation}
b_0 = 0,   \hskip 10mm  b_{k+1} = b_k + 2^{-N/2} a'_k  ~~~ (k = 0, 1, ..., 2^N-1) ,
\end{equation}
with $2^N$ independent normally distributed random variables $a'_k \in
\N(0,1)$.  Similar computation relying on wavelet representations
requires one random variable for a linear shift and $2^k$ random
variables at each scale $k$, $k$ ranging from $0$ to $N-1$.  The total
number is then $1 + (1 + 2 + 4 + ... + 2^{N-1}) = 2^N$.  It is not
surprising that both schemes require the same degree of randomness to
represent a Brownian path at chosen scale.  The wavelet representation
does not reduce the complexity or randomness, but re-organize the data
in a hierarchical structure to facilitate their use.  For instance,
formula (\ref{eq:Bomega1}) accesses approximate positions of Brownian
motion at any time point $t$, not necessarily $t_k$.  In a classical
scheme, one could use a linear interpolation between two neighboring
points to get the same result.  Again, the wavelets do not bring new
features which are not available by conventional techniques, but
provide another, structured and efficient, representation.

Throughout the above sections, Brownian motion was constructed on the
unit interval for convenience.  The constructed process can be easily
rescaled to any finite interval, while an extension to $\R_+$ or $\R$
is possible as well.  Finally, an extension to isotropic Brownian
motion in $\R^d$ is obtained by taking $d$ independent samples of
one-dimensional Brownian motion.

\section{Beyond Brownian motion}

\subsection{Fractional Brownian motion}

A similar technique can be applied to construct and study fractional
Brownian motion which is also known as random fractal velocity field
\cite{Kolmogorov40,Mandelbrot68}.  For instance, random fractal
velocity field with the Hurst exponent $H=1/3$ (defined below)
corresponds to the Kolmogorov spectrum in high Reynolds number
turbulence \cite{Mandelbrot,Lesieur,McComb,Majda99}.  Fractional
Brownian motion, as a Gaussian stochastic process with long-range
correlations, has found numerous applications in different fields,
ranging from transport phenomena in porous media
\cite{Elliott94,Elliott95,Elliott95b,Elliott97,Bunde} to analysis of
financial markets \cite{Mandelbrot97}.

P. L\'evy proposed the first extension of Eq.~(\ref{eq:Bnoise}) by
using the Riemann-Liouville fractional integration which can be
thought of as a moving average of a Gaussian white noise \cite{Levy53}
\begin{equation}
\label{eq:fBM}
\tilde{B}_H(t) = \frac{1}{\Gamma(H+\frac12)}\int\limits_{0}^t (t-s)^{H-\frac12} dW(s) ,
\end{equation}
where $0 < H < 1$ is the Hurst exponent, and $\Gamma(H+1/2)$ is the
normalization factor ($\Gamma(z)$ being the Gamma function).
Mandelbrot and van Ness discussed the limitations of this definition
(e.g., its strong emphasis on the origin) and proposed to use the Weyl
fractional integral that yields \cite{Mandelbrot68}
\begin{equation}
\label{eq:fBM2}
\begin{split}
B_H(t) & = \frac{1}{\Gamma(H+\frac12)} \biggl\{\int\limits_{-\infty}^0 \bigl[(t-s)^{H-\frac12} - (-s)^{H-\frac12}\bigr] dW(s)  \\
& + \int\limits_{0}^t (t-s)^{H-\frac12} dW(s) \biggr\}  \\
\end{split}
\end{equation}
for $t > 0$ (and similar for $t < 0$).  The last representation can
also be written as (see \cite{Decreusefond99})
\begin{equation}
\label{eq:fBM3}
B_H(t) = \int\limits_{0}^t K_H(t,s)~ dW(s) , 
\end{equation}
where
\begin{equation}
K_H(t,s) = \frac{(t-s)^{H-\frac12}}{\Gamma(H+\frac12)} ~ _2F_1\biggl(H-\frac12, \frac12-H; H+\frac12; 1-\frac{t}{s}\biggr) ,
\end{equation}
and $_2F_1(a,b;c;z)$ is the hypergeometric function.  The ordinary
Brownian motion is retrieved at $H = 1/2$ for all cases.

Using a wavelet representation (\ref{eq:dW_psi}) for the Gaussian
white noise, one obtains
\begin{equation}
\label{eq:fBM4}
B_H(t) = \sum\limits_{i=0}^\infty \hat{a}_{i} \int\limits_0^t ds~K_H(t,s) \psi_i(s) .
\end{equation}
The integrals can be evaluated using an appropriate basis
$\{\psi_i(t)\}$.  Moreover, the moment cancellation property
(\ref{eq:moments}) for the Alpert-Rokhlin multiwavelets with a large
enough order $q$ guarantees that the integrals in Eq.~(\ref{eq:fBM4})
are highly localized, yielding a rapid convergence of the above sum.
This convergence is a key point for efficient numerical algorithms for
simulation of fractional Brownian motion (see
\cite{Elliott94,Elliott95,Elliott95b,Elliott97}).  Among other
numerical methods, we mention alternative wavelet representations
\cite{Wornell90,Flandrin92,Sellan95,Arby96} (e.g., the method by Arby
and Sellan is implemented in the Matlab function `wfmb'), circulant
embedding of the covariance matrix \cite{Davies87,Wood94,Dietrich97},
and random midpoint displacement method \cite{Fournier82} which is
often used in computer graphics to generate random two-dimensional
landscapes.

In general, the kernel $K_H(t,t')$ can be replaced by any convenient
kernel to extend this approach to various Gaussian processes.  For
instance, setting $K(t,t') = e^{-\theta(t-t')}$ yields the
Ornstein-Uhlenbeck process \cite{Uhlenbeck30,Risken}.  Similarly, one
can deal with various stochastic dynamics generated by Langevin
equations \cite{Coffey}.

\subsection{Gaussian Free Field and its extensions}

Brownian motion is the integral of the Gaussian white noise which, in
turn, is obtained as a linear combination of orthonormal functions
$\{\psi_i\}$ forming a complete basis of the space $L^2([0,1])$, with
standard Gaussian weights.  This construction can be extended to any
separable Hilbert space $H$.  However, whatever the functional space
$H$ is taken, a linear combination of its orthonormal basis functions
with standard Gaussian weights does not belong to this space (the
argument is the same as for $L^2$ space, the norm of such a linear
combination being infinite).  In particular, the Gaussian white noise
is not a function but a distribution.  In order to construct
extensions of the Gaussian white noise with desired properties, one
needs to carefully choose the Hilbert space $H$.  In this section, we
briefly discuss two such extensions: Gaussian free field (GFF)
\cite{Sheffield07} and fractional Gaussian field (FGF) with
logarithmic correlations \cite{Lodhia14,Duplantier14}.  The GFF
appears as the basic description of massless non-interacting particles
in field theories.  Both GFF and FGF constitute important models in
different areas of physics, from astrophysics (describing stochastic
anisotropy in cosmic microwave background) to critical phenomena,
quantum physics, and turbulence
\cite{Bardeen86,Kobayashi11,Fernandez,Dodelson}.  While the ``sequence'' of
random variables of Brownian motion $B(t)$ was naturally parameterized
by ``time'' $t$ (a real number from the unit interval or, in general,
from $\R$), random variables of a field can in general be
parameterized by points from an Euclidean domain, a manifold, or a
graph.  For instance, one can speak about random surfaces which can
model landscapes (e.g., mountains) or ocean's water surface, in which
the height is parameterized by two coordinates.  The geometrical
structure of ``smooth'' random surfaces was thoroughly investigated
\cite{Adler,Adler2,Vanmarcke}, especially for Gaussian fields which
are fully characterized by their mean and covariance.  Important
examples of smooth Gaussian fields are the random plane waves and
random spherical harmonics (see \cite{Berry77,Nazarov10}).  In turn,
the GFF and FGF are examples of highly irregular random fields.  As
Brownian motion can be obtained as the limit of discrete random walks,
the Gaussian free field in two dimensions appears in the limit of
discrete random surfaces \cite{Kondev95,Schramm09}.

The Gaussian free field is constructed by choosing the Dirichlet
Hilbert space $H_\nabla(\Omega)$, in which the scalar product of two
functions $f$ and $g$ is defined as
\begin{equation}
\label{eq:Hilbert}
\langle f,g\rangle_{H_\nabla(\Omega)} = \int\limits_\Omega dx ~ (\nabla f \cdot \nabla g)
\end{equation}
for a given Euclidean domain $\Omega\subset \R^d$.  When $\Omega$ is
bounded, an orthonormal basis of this space can be obtained by setting
$\psi_i(x) = \lambda_i^{-1/2} u_i(x)$, where $u_i(x)$ are the
$L^2$-normalized Dirichlet eigenfunctions of the Laplace operator:
$\Delta u_i(x) + \lambda_i u_i(x) = 0$ in $\Omega$ with $u_i(x) = 0$
at the boundary $\partial\Omega$, and $\lambda_i$ are the
corresponding eigenvalues.  The Gaussian free field on $\Omega$ is
defined as
\begin{equation}
\label{eq:GFF_2d}
F(x) = \sum\limits_{i=0}^\infty a_i~ \psi_i(x) 
\end{equation}
where $a_i\in\N(0,1)$ are independent Gaussian weights.  Given that
the eigenvalues $\lambda_i$ asymptotically grow as $i^{2/d}$ according
to the Weyl's law \cite{Courant,Grebenkov13}, the sum in
Eq. (\ref{eq:GFF_2d}) is convergent for $d=1$ (in which case the GFF
is simply a Brownian bridge) but diverges in higher dimensions.  In
the plane, this sum barely misses the convergence, being
logarithmically diverging.  In quantum field theory, it is related to
the infra-red divergence for massless particles.  As a consequence,
$F(x)$ is not a function but a distribution for $d > 1$.  Being a
linear combination of normal variables, the field $F(x)$ is Gaussian
and thus is fully characterized by its mean $\E\{F(x)\} = 0$ and the
covariance
\begin{equation}
\label{eq:GFF_covar}
\E\{F(x_1) F(x_2)\} = \sum\limits_{i=0}^\infty \lambda_i^{-1} u_i(x_1) u_i(x_2) = G(x_1,x_2),
\end{equation}
where the right-hand side can be recognized as the Green function
$G(x_1,x_2)$ of the Laplace operator in the domain $\Omega$.
Alternatively, one could define the GFF by setting the covariance
equal to the Green function (in which case the definition holds even
for unbounded domains).  Strictly speaking, since the sum in
Eq. (\ref{eq:GFF_2d}) diverges for $d>1$, the GFF should be treated as
a distribution by its action on every fixed test function $\phi(x)$.
In particular, the covariance should be given by covariance of actions
of $F$ on two test functions $\phi_1$ and $\phi_2$:
\begin{equation}
\E\{F \phi(x_1),  F \phi(x_2)\} = \hspace*{-1mm} \int\limits_{\Omega\times \Omega} \hspace*{-1mm} dx_1 dx_2 ~ G(x_1-x_2) ~ \phi_1(x_1)~ \phi_2(x_2) .
\end{equation}

The fractional Gaussian fields can be obtained by replacing the
gradient operators $\nabla$ in the scalar product (\ref{eq:Hilbert})
by fractional Laplacians \cite{Chen10},
\begin{equation}
\label{eq:Hilbert_nu}
\langle f,g\rangle_{H_\nabla^\nu(\Omega)} = \int\limits_\Omega dx ~ ((-\Delta)^\nu f \cdot (-\Delta)^\nu g),
\end{equation}
for a positive $\nu$.  The Dirichlet Hilbert space is retrieved for
$\nu = 1/2$.  In a similar way, the functions $\psi_i(x) =
\lambda_i^{-\nu/2} u_i(x)$ form a basis of the space
$H_\nabla^\nu(\Omega)$, from which the FGF is constructed as in
Eq. (\ref{eq:GFF_2d}).  This sum is convergent for $\nu > d/4$ and
divergent otherwise.  In the particular case $\nu = d/4$, the FGF is
logarithmically divergent in all dimensions.  Note that the FGF
coincides with the GFF in the plane ($d = 2$).  In particular, the FGF
with logarithmic correlations on the unit interval reads explicitly as
\begin{equation}
F(t) = \sqrt{\frac{2}{\pi}} \sum\limits_{k=1}^\infty  a_k ~\frac{\sin(\pi k t)}{\sqrt{k}}.
\end{equation}
More generally, one can take any orthonormal basis $\{u_i(t)\}$ of
$L^2([0,1])$ and then apply the operator $(-\Delta)^{-1/4} =
\bigl(\frac{d}{dt}\bigr)^{-1/2}$ to get the basis of
$H_{\nabla}^{1/4}([0,1])$.  The advantage of the Laplacian eigenbasis
is that the fractional integral operator
$\bigl(\frac{d}{dt}\bigr)^{-1/2}$ (which can be defined through the
Fourier transform) is replaced by multiplication by
$\lambda_i^{-1/4}$.  As a consequence, the FGF with logarithmic
corrections in one dimension appears as the half-derivative of
Brownian motion:
\begin{equation}
\label{eq:auxil5}
\begin{split}
F(t) & = \sum\limits_{i=0}^\infty a_i  \biggl(\frac{d}{dt}\biggr)^{-1/2} u_i(t) \\
& = \sum\limits_{i=0}^\infty a_i  \biggl(\frac{d}{dt}\biggr)^{1/2} \int\limits_0^t dt' ~ u_i(t')
= \biggl(\frac{d}{dt}\biggr)^{1/2} B(t). \\
\end{split}
\end{equation}
This formula reveals a very close relation between these two processes
which are often considered as distinct objects.  Note that the series
in Eq. (\ref{eq:auxil5}) diverges logarithmically, while the
derivative of order $1/2-\epsilon$ would lead to converging series.
In other words, Brownian motion belongs to the H\"older space
$H_{1/2-\epsilon}$ (for any $\epsilon > 0$) so that its derivatives of
order less than $1/2$ exist, but $1/2$ and higher do not.  As for
Brownian motion, the explicit closed formula (\ref{eq:GFF_2d}) allows
one to sample random realizations of the FGF by picking up $\omega$
from the unit interval, i.e., the probability space for this process
is nothing else than the unit interval with uniform measure.

\section{Restricted diffusion}

In this section, we briefly discuss how multiscaling and dyadic
decompositions help simulating restricted diffusion.  This is an
ubiquitous problem in physics (e.g., transport in porous media),
chemistry (e.g., heterogeneous catalysis), biology and physiology
(e.g., diffusion in cells, tissues, and organs).  When Brownian motion
is restricted, physico-chemical or biological interactions between the
diffusing particle and the interface of a confining medium should be
taken into account.  For instance, paramagnetic impurities dispersed
on a liquid/solid interface cause surface relaxation in nuclear
magnetic resonance (NMR) experiments \cite{Callaghan,Grebenkov07};
cellular membranes allow for a semi-permeable transport through the
boundary \cite{Alberts,Bressloff13,Benichou14}; chemical reaction may
transform the particle or alter its diffusive properties
\cite{Wilemski73,Coppens99}.  While an accurate description of
these processes at the microscopic level is challenging, the contact
with the interface is very rapid at macroscopic scales, allowing one
to resort to an effective description of the surface transport by an
absorption/reflection mechanism \cite{Sapoval94,Sapoval96}.  This
mathematical process is known as (partially) reflected Brownian motion
\cite{Grebenkov06,Grebenkov06c,Singer08}.

The presence of a boundary drastically changes the properties of
Brownian motion (e.g., the reflecting boundary forces the process to
remain inside the domain) so that earlier wavelet representations
cannot directly incorporate the effect of the boundary.  Since
restricted diffusion is relevant for most physical, chemical and
biological applications, various Monte Carlo methods have been
developed for simulating this stochastic process, computing the
related statistics (e.g., the first passage times \cite{Redner}), and
solving the underlying boundary value problems
\cite{Sabelfeld,Sabelfeld2,Milstein,Grebenkov14}.  The slow
convergence of Monte Carlo techniques (typically of the order of
$1/\sqrt{M}$ in the number of trials) requires fast generation of
Brownian paths.  The simplest generation of a Brownian path at
successive times $\delta, 2\delta, 3\delta, \ldots$ by adding normally
distributed displacements and checking the boundary effects at each
step becomes inefficient in multiscale media.  In fact, tiny
geometrical details of the medium require the use of comparably small
displacements, resulting in a very large number of steps needed to
model large-scale excursions.

To overcome this limitation, the concept of fast random walks was
proposed \cite{Muller56}.  The basic idea consists in adapting
displacements to the local geometrical environment, performing as
large as possible displacements without violating the properties of
Brownian motion.  When the walker is at point $x$, the largest
displacement is possible at the distance $|x-\pa|$ between $x$ and the
boundary $\pa$ of an Euclidean domain $\Omega\subset \R^d$.  In fact,
the ball $B(x,|x-\pa|)$ of radius $|x-\pa|$ does not contain any
``obstacle'' (e.g., piece of boundary) to the walker.  Since Brownian
motion is continuous, it must leave the ball before approaching the
boundary of the confining domain.  The rotation symmetry implies that
the exit points are distributed uniformly over the boundary of the
ball.  Instead of modeling the fully-resolved trajectory of Brownian
motion inside the ball, one can just pick up at random a point $x'$ on
the sphere of radius $|x-\pa|$ and move the random walker at this new
position.  The random duration of this displacement can be easily
generated \cite{Redner,Grebenkov14}.  From here, one draws a new ball
$B(x',|x'-\pa|)$, and so on, until the walker approaches the boundary
$\pa$ closer than a chosen threshold.  From this point, an appropriate
boundary effect (e.g., absorption, relaxation, chemical
transformation, permeation, reflection, etc.) is implemented.  Due to
its efficiency, fast random walk algorithms have been used to simulate
diffusion-limited aggregates (DLA)\cite{Meakin85,Ossadnik91}, to
generate the harmonic measure on fractals
\cite{Grebenkov05,Grebenkov05b,Grebenkov06c}, to model
diffusion-reaction phenomena in spherical packs
\cite{Torquato89,Zheng89}, to compute the signal attenuation in
pulsed-gradient spin-echo experiments \cite{Leibig93,Grebenkov11},
etc.  In this section, we focus on multiscale tools to estimate the
distance, while other aspects of fast random walk algorithms can be
found elsewhere \cite{Grebenkov14}.

\subsection{Distance to a boundary}

The efficiency of fast random walk algorithms fully relies on the
ability to rapidly estimate the distance between any point (e.g., the
current position of the walker) and the boundary.  Multiscale dyadic
decompositions provide an efficient way to these estimates.  To
illustrate the idea, we first consider the one-dimensional case and
then discuss its straightforward extension to the multidimensional
case.

In one dimension, the problem can be formulated as follows: given a
set $\{x_n\}$ of $N$ ``boundary'' points on the unit interval, how the
distance to this set from another point $x$ can be estimated in a
rapid way?  Successive computation of the distances $|x-x_n|$ and
finding their minimum is of course the simplest but the slowest way
(of order of $N$).  Instead of computing the distances to all boundary
points, one can split the unit interval into two half-intervals, and
check the distance to the points belonging to the half-interval that
contains $x$.  If the distribution of points $x_n$ on $[0,1]$ is more
or less uniform, this division approximately halves the number of
computations.  In the same spirit, splitting on subintervals of length
$1/4$, $1/8$, etc. would reduce the number of computations roughly by
factors $4$, $8$, etc.  Using such dyadic decompositions, one needs
approximately $\log_2(N)$ splitting to attend the level when one (or
few) point $x_n$ belongs to the same subinterval as $x$.  The number
of computations is then of order of $\log_2(N)$ (assuming the
distribution of boundary points is more or less uniform).  Moreover,
if computation is carried with a desired precision $\ve$ (to consider
$x_n$ as ``pointlike'', one needs $\ve \ll 1/N$), one can continue
splitting up to the level $\log_2(1/\ve)$ so that the length of
subintervals becomes smaller than $\ve$.  Since the points $\{x_n\}$
are stored with precision $\ve$, one cannot distinguish two points at
any scale smaller than $\ve$.  Consequently, any subinterval of length
$\ve$ can be either vacant, or occupied by only one point $x_n$ (two
points from the same subinterval would be indistinguishable).  In this
case, the number of computations, $\log_2(1/\ve)$, is actually
independent of whether the distribution of points $x_n$ is uniform or
not.  In other words, this algorithm can be applied for any finite set
of points $x_n$ that are all distinguishable at scale $\ve$, i.e.,
$|x_n - x_m| \geq \ve$ for any $n$ and $m$.

\subsection{Dyadic decomposition}  

\begin{figure}
\begin{center}
\includegraphics[width=85mm]{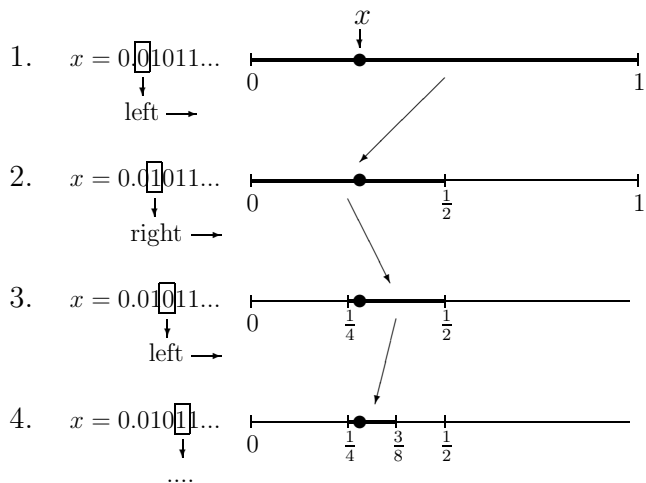} 
\end{center}
\caption{
Construction of a dyadic tree of subintervals for a given boundary
point $x$ by its binary expansion. }
\label{fig:dyadic}
\end{figure}

\begin{figure}
\begin{center}
\includegraphics[width=42mm]{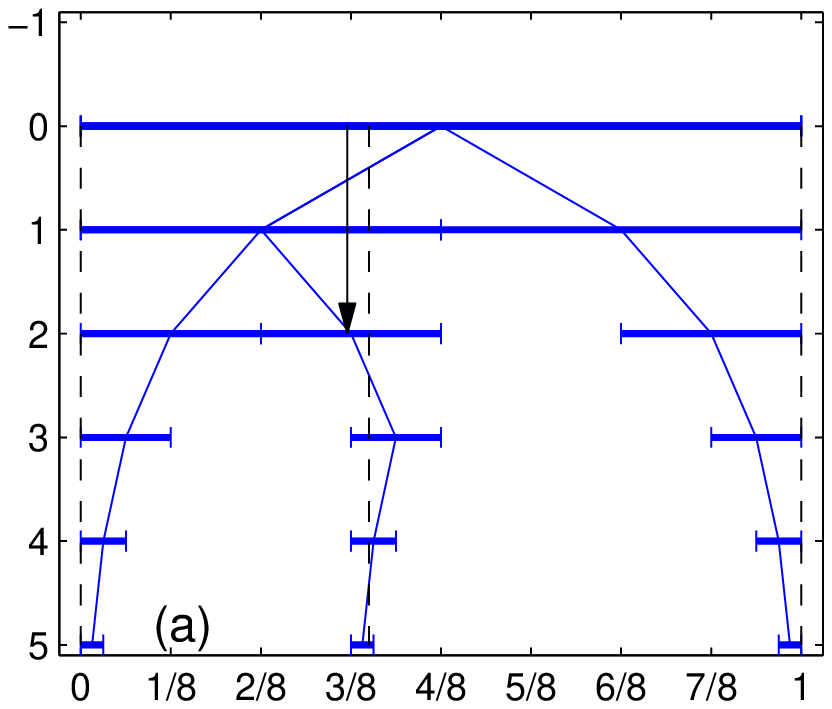} 
\vskip 3mm
\includegraphics[width=84mm]{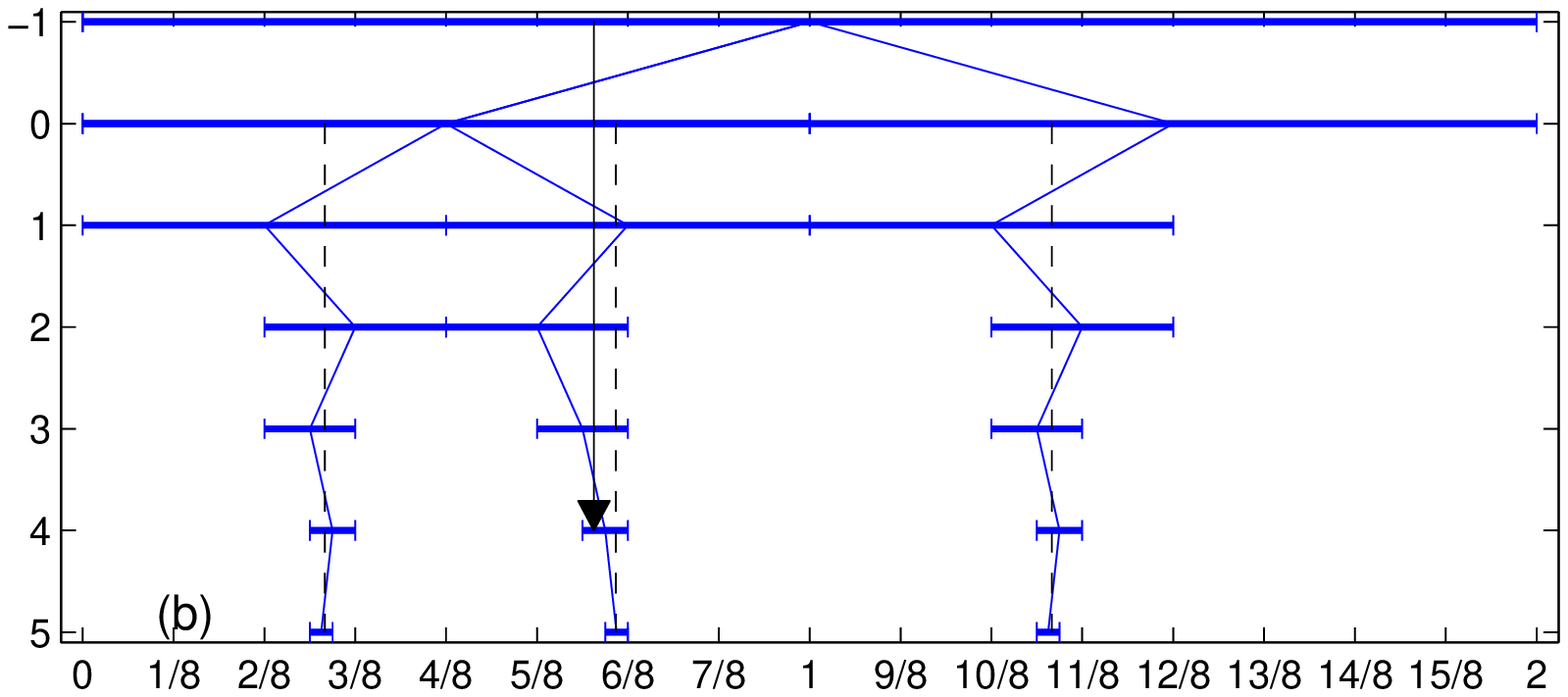} 
\end{center}
\caption{
{\bf (a)} Example of a dyadic decomposition of the unit interval with
three boundary points $\{ 0,~ 0.4,~ 1\}$ (shown by vertical dashed
lines) and the related tree of subintervals at five levels ($\ve =
2^{-5}$).  For a test point $x = 0.37$ (shown by arrow), one uses its
binary expansion $x = 0.01011\ldots$ to navigate over the tree.  The
descend is stopped at level $n = 2$ since $x\notin I_{33} =
[3/8,4/8]$.  However, the rough estimate, $2^{-n-1} = 0.125$, of the
distance between $x$ and the boundary, $|x - \pa| = |0.37-0.4| =
0.03$, obviously fails.  To apply the $1/3$-trick, one constructs the
dyadic decomposition for the boundary points shifted by $1/3$ {\bf
(b)}.  In this tree, the descent for the point $x + 1/3$ is stopped at
level $n' = 4$.  The combined lower estimate $2^{-\max\{n,n'\}}/6 =
2^{-4}/6 \approx 0.0104$ is valid.}
\label{fig:1D_decomp}
\end{figure}

For practical implementation, the boundary points $x_n$ are used to
generate a dyadic tree of subintervals at the scales ranging from $1$
to $\log_2(1/\ve)$.  In turn, the binary expansion of the test point
$x$ is used to ``navigate'' search on the tree
(Fig. \ref{fig:dyadic}).  In fact, one can associate to a given point
$x\in [0,1]$ a sequence of dyadic intervals $I_{n,\lfloor 2^n
x\rfloor}$ such that $x\in I_{n,\lfloor 2^n x\rfloor}$ at any scale
$n$.  At each scale $n$, one chooses the left or the right subinterval
depending on whether the $n^{\rm th}$ bit is $0$ or $1$.  Applying
this procedure to all boundary points $x_n$, one can generate a dyadic
decomposition of the boundary.  Figure~\ref{fig:1D_decomp}a shows an
example with three boundary points $\{0,~ 0.4,~ 1\}$ at five scales,
from $2^0$ to the smallest one $\ve = 2^{-5}$.  The dyadic
decomposition is stored as a tree, where a vertex is associated with a
subinterval.  Each vertex can be connected to one, two, or three other
vertices (see Fig.~\ref{fig:1D_decomp}a).  The ``height'' of the
vertex from the ``root'' determines the scale of the corresponding
subinterval.

Once a dyadic decomposition for a given boundary is constructed, it
can be used to estimate the distance to the boundary from any point
$x$.  In fact, one can easily (and very rapidly) find the smallest
subinterval $I_{n,\lfloor 2^n x\rfloor}$ containing simultaneously $x$
and some boundary point.  For this purpose, one starts from the
``root'' vertex and descends on the tree using the bits of $x$ to
choose left or right edges at each scale.  The descend is stopped when
there is no edge to follow (Fig. \ref{fig:1D_decomp}a).  Once the
smallest common interval $I_{n,\lfloor 2^n x\rfloor}$ is found, there
are two option: either $n = \lfloor \log_2(1/\ve) \rfloor$ so that the
point $x$ is indistinguishable from some boundary point at scale
$\ve$, and the distance estimate is set to $0$; or $n < \lfloor
\log_2(1/\ve) \rfloor$, and the distance to the boundary can be
roughly estimated as $2^{-n-1}$ since the next subdivision must
separate the point $x$ from the boundary points.  However, this
simplistic argument fails when the point $x$ and the closest boundary
point lie on opposite sides of the midpoint of the smallest common
interval $I_{n,\lfloor 2^n x\rfloor}$, but very close to each other.
Although the next subdivision separates these two points, the distance
between them can be arbitrarily small so that the rough estimate
$2^{-n-1}$ is wrong, as illustrated on Fig. \ref{fig:1D_decomp}a.  To
get the correct lower estimate, one can apply the so-called
``1/3-trick''.

\subsection{The 1/3-trick}

The $1/3$-trick can be easily illustrated for two points $x$ and $y$.
Let $I_{n,\lfloor 2^nx\rfloor}$ be the smallest common interval
containing both points $x$ and $y$.  Suppose that the distance between
these points is smaller than $2^{-n}/6$.  We consider the points
$x'=x+1/3$ and $y'=y+1/3$ (shifted by $1/3$), and determine their
smallest common interval $I_{n',\lfloor 2^{n'} x'\rfloor}$.  As shown
in Appendix \ref{sec:trick_proof}, the distance between the points
$x'$ and $y'$ (and thus between the points $x$ and $y$) is larger than
$2^{-n'}/6$.  It is thus sufficient to find the smallest common
intervals for the pair $x,y$ and its shifted counterpart $x',y'$, and
the distance between the points is bounded below as
\begin{equation*}
|x - y| = |x' - y'| \geq 2^{-\max\{n,n'\}}/6 .
\end{equation*}

This simple fact allows one to rapidly estimate the distance to
boundary points.  Let us consider a new boundary $\pa'$ which is
obtained by shifting the old one by $1/3$: $\pa' = \{ x\in\R ~:~
x-1/3\in \pa\}$.  For the new boundary, another dyadic tree of
subintervals can be constructed (Fig. \ref{fig:1D_decomp}b) in order
to estimate the distance between $\pa'$ and the shifted point $x' =
x+1/3$ at a given scale $\ve$.  While this construction may appear
redundant at first thought because $|x-\pa| = |x'-\pa'|$, the crucial
point is that we search for a lower estimate at a finite scale at
which two dyadic trees are different.  Performing the descend over
both dyadic trees (using the binary expasion of $x$ and $x'$,
respectively), one identifies the level $n$ (resp. $n'$) of the
smallest common interval of $x$ (resp. $x'$) and the closest boundary
point (resp. shifted closest boundary point).  The lower estimate of
the distance is then
\begin{equation*}
|x - \pa| = |x' - \pa'| \geq 2^{-\max\{n,n'\}}/6 .
\end{equation*}

\subsection{Higher dimensions}

Similar constructions are applicable in higher dimensions which are
more relevant for applications.  The lower estimate relies on the
generalized mean inequality:
\begin{equation}
\label{eq:generalized_mean}
M_q(c_1,\ldots,c_n) \leq M_p(c_1,\ldots,c_n)  \qquad (q < p),
\end{equation}
where the generalized mean $M_p(c_1,\ldots,c_n)$ of $n$ positive
numbers $c_1,\ldots, c_n$ is 
\begin{equation}
M_p(c_1,\ldots,c_n) = \left(\frac{1}{n}\sum\limits_{k=1}^n c_k^p\right)^{1/p} .
\end{equation}
Setting $q = 1$, $p = 2$, and $c_k = |x_k - y_k|$ for two points
$x=(x_1,\ldots,x_d)$ and $y=(y_1,\ldots,y_d)$,
Eq. (\ref{eq:generalized_mean}) yields the lower estimate of their
Euclidean distance:
\begin{equation}
\label{eq:dist_d}
|x-y| = \left(\sum\limits_{k=1}^d (x_k-y_k)^2\right)^{1/2} \geq  \frac{1}{\sqrt{d}}\sum\limits_{k=1}^d |x_k-y_k| . 
\end{equation}
Constructing two dyadic trees for each coordinate as earlier, one can
then estimate each term $|x_k-y_k|$ and thus the distance from any
point $x = (x_1,\ldots,x_d)$ to the boundary.  In high dimensions ($d
\gg 1$), the right-hand side of the inequality (\ref{eq:dist_d}) is
strongly attenuated by the prefactor $1/\sqrt{d}$.  This is an example
of the so-called ``curse of dimensionality''.  To overcome this
difficulty, one can implement random rotations and translations of the
boundary.  Although these transformations preserve the distance, they
can improve the lower bound at a finite scale.  Note that other
multiscale constructions and the related searchable data structures
can also be used such as Whitney decompositions of the computational
domain (in which the size of each square (or cube) paving the domain
is comparable to the distance to the boundary), quadtrees (or
Q-trees), k-d trees, etc. \cite{deBerg}.

\subsection{Overall efficiency}

The advantages of multiscale dyadic trees are numerous: the simplicity
of construction, the generality of boundary shapes, the rapidity of
distance estimation, the flexibility for shape modifications, and low
memory usage.  In fact, for a given precision $\ve$, the storage of
the dyadic tree requires at worst $N\log_2(1/\ve)$ intervals (i.e.,
$\log_2(1/\ve)$ levels for each boundary point, for $N$ points).  In
practice, this number is much smaller since many boundary points share
the same interval at larger scales (e.g., the interval of size $1$ is
shared by all boundary points).

The crucial point is that geometrical structure of the boundary does
not matter at all: the method works for a random Cantor dust as well
as for a circle.  Moreover, the tree-like representation is highly
adaptive, allowing one to modify the boundary from one set of
simulations to other (or even from one run to the other).  This
feature can be very useful to study diffusion-controlled growth
processes like DLA or transport phenomena in domains with moving
boundaries.

\section{Conclusions}

In this paper, we revised the multiscale construction of Gaussian
processes and fields.  First, the Haar wavelet representation of
Brownian motion was explicitly constructed as a natural way to refine
the geometrical features of a Brownian path under magnification.
Since the Haar functions form a complete basis in the space
$L^2([0,1])$ and their weights are Gaussian, such a representation can
be extended to any complete basis $\{\psi_i(t)\}$ of $L^2([0,1])$,
wavelet-like or not.  In other words, the construction of Brownian
motion has two separate ``ingredients'': deterministic functions
$\psi_i(t)$ capturing geometrical details, and their random weights
$\hat{a}_i$.  The choice of the functions $\psi_i(t)$ is voluntary
that gives a certain freedom and flexibility in dealing with different
problems.  Qualitatively, this choice determines the way of connecting
successive positions of Brownian motion at a given scale.  On the
opposite, the random weights determine the intrinsic stochastic
properties of Brownian motion, independently of our choice of the
basis $\{\psi_i(t)\}$.

The multiscale construction gives not only a simple closed formula for
Brownian motion, but reveals its fundamental properties.  For
instance, continuity and non-differentiability of Brownian paths
naturally follow from this construction.  In addition, we discussed a
closed mapping from the sampling unit interval onto the space of
Brownian paths.  Sampling Brownian paths can therefore be formally
reduced to picking up a real number $\omega$ with the uniform measure.
This construction does not require elaborate notions from modern
probability theory such as Wiener measures, sigma-algebras,
filtrations, etc.  Although these notions are useful, the explicit
multiscale construction is much easier for non-mathematicians.

These concepts are not limited to Brownian motion.  We illustrated how
fractional Brownian motion, Gaussian free field, and fractional
Gaussian fields can be constructed in a very similar way.  Extensions
to other Gaussian processes were also mentioned.  Finally, we briefly
discussed how the multiscale concepts can be used for simulating
restricted diffusion (i.e., Brownian motion in confining domains).
The dyadic subdivision and the related hierarchical (multiscale) tree
of subintervals allow one to rapidly estimate the distance to the
boundary and thus to generate random displacements adapted to the
local geometrical environment.  Such fast random walk algorithms have
found numerous applications.  As for usual Brownian motion, dyadic
decompositions appear as natural tools to store and rapidly access the
geometrical information, resulting in fast algorithms.

\section*{Acknowledgments}

DG acknowledges the financial support by ANR Project
ANR-13-JSV5-0006-01.  DB was partially funded by EPSRC Fellowship ref.
EP/M002896/1.

\appendix
\section{Conditional law}
\label{sec:cond_law}

We explain why the distribution of the position of Brownian motion
$y=B(1/2)$ at time $t=1/2$ under conditions $B(0)=0$ and $B(1) = x$ is
given by the normal law $\N(x/2,1/4)$.

Since the increments of Brownian motion on the unit intervals
$(0,1/2)$ and $(1/2,1)$ are independent, the joint probability density
$p\{ B(1/2) = y,~ B(1) = x \}$ is simply equal to the product of the
probability density $p\{ B(1/2)-B(0) = y\}$ to have the first
increment equal to $y$ and the probability density $p\{ B(1) - B(1/2)
= x-y \}$ to have the second increment equal to $x-y$.  These
densities are given by normal laws with variance $1/2$, yielding
\begin{equation*}
\begin{split}
p\{ B(1/2) = y,~ B(1) = x \} & = \left(\frac{e^{-\frac{y^2/2}{1/2}}}{\sqrt{2\pi}\sqrt{1/2}}\right)
\left(\frac{e^{-\frac{(x-y)^2/2}{1/2}}}{\sqrt{2\pi}\sqrt{1/2}}\right) \\
& = \left(\frac{e^{-\frac{(y-x/2)^2/2}{1/4}}}{\sqrt{2\pi}\sqrt{1/4}}\right) \left(\frac{e^{-x^2/2}}{\sqrt{2\pi}}\right) . \\
\end{split}
\end{equation*}
The second factor is simply the probability density for $x = B(1)$,
while the first factor is the conditional probability density we are
looking for:
\begin{equation*}
p\{ B(1/2) = y~|~ B(1) = x \} = \frac{e^{-\frac{(y-x/2)^2/2}{1/4}}}{\sqrt{2\pi}~\sqrt{1/4}} .
\end{equation*}
This is the normal distribution with the mean $x/2$ and the variance
$1/4$.

\section{Continuity and non-differentiability of Brownian motion}
\label{sec:continuity}

In this section, we illustrate how the wavelet representation of
Brownian motion can be used to prove its basic properties such as
continuity and non-differentiability.

\subsection{Continuity and convergence}

First, we show that a partial sum approximation converges to Brownian
motion in both $L^\infty$ and $L^2$ norms.  Let us denote
\begin{equation}
F_n(t) = \sum\limits_{k=0}^{2^n-1} a_{nk} h_{nk}(t)
\end{equation}
so that $B(t) = a_0 t + F_0(t) + F_1(t) + \ldots$ according to
Eq. (\ref{eq:Bomega1}).  Each function $F_n$ is a sum of hat functions
with disjoint supports.  To estimate the size of $F_n$ we need some
upper bound on $a_{nk}$.  We are going to show that $|a_{nk}|<n$ for
all sufficiently large $n$.

Since $a_{nk}\in \N(0,1)$ are standard normal variables, we observe that
\begin{equation*}
\P(|a_{nk}|\ge n) = 2\int\limits_n^\infty dz ~ \frac{e^{-z^2/2}}{\sqrt{2\pi}} \le  e^{-n^2/2}
\end{equation*}
for sufficiently large $n$.  This inequality yields
\begin{equation*}
\sum\limits_{n=0}^\infty \sum\limits_{k=0}^{2^n-1} \P(|a_{nk}|\ge n)\le \sum\limits_{n=0}^{\infty} 2^{n}e^{-n^2/2}<\infty.
\end{equation*}
By Borel-Cantelli lemma this implies that $|a_{nk}|<n$ for all but
finitely many coefficients $a_{nk}$.  In other words, almost surely,
there is finite, but random, $N^*$ such that $|a_{nk}|<n$ if $n>N^*$.
For such $n$ we have
\begin{equation*}
\|F_n\|_{L^\infty([0,1])} = 2^{-n/2-1}\max_k \{|a_{nk}|\} < n 2^{-n/2-1}.  
\end{equation*}
Since the sum of these norms converges, the series $a_0 t + \sum_n
F_n(t)$ converges to $B(t)$ in $L^\infty$ norm (uniformly), i.e. for
any $\epsilon > 0$
\begin{equation}
\label{eq:Bconv1}
\lim\limits_{N\to\infty} \P\left\{ \left\| B(t) - B_N(t)\right\|_{L^\infty([0,1])} > \epsilon \right\} = 0 ,
\end{equation}
where
\begin{equation}
B_N(t) = a_0 t + \sum\limits_{n=0}^N \sum\limits_{k=0}^{2^n-1} a_{nk} h_{nk}(t)
\end{equation}
is a partial sum approximation of Brownian motion at scale $2^{-N}$.
This proves that Brownian motion is almost surely continuous.
Moreover, the remainder of a partial sum approximation at scale
$2^{-N}$ is exponentially small:
\begin{equation}
\label{eq:approx_remainder}
\left\| B(t) - B_N(t) \right\|_{L^\infty} 
< \sum\limits_{n=N+1}^\infty \left\| F_n(t) \right\|_{L^\infty} < \frac{3(N+3)}{2^{1+N/2}} ,
\end{equation}
when $N$ is large enough.

The $L^2$ convergence can be shown in the same way.  We have
\begin{equation*}
\|F_n\|^2_{L^2([0,1])} = \sum\limits_{k=0}^{2^n-1} |a_{nk}|^2 \|h_{nk}\|^2_{L^2([0,1])} 
< 2^n ~ n^2 ~ 2^{-2n},  
\end{equation*}
where we used $|a_{nk}| < n$ for large enough $n$.  This inequality
implies the convergence of $\sum_n \|F_n\|_{L^2([0,1])}$ with
probability one and hence the series in Eq. (\ref{eq:Bomega1})
converges almost surely in $L^2$ norm:
\begin{equation}
\label{eq:Bconv2}
\lim\limits_{N\to\infty} \E\left\{ \left\| B(t) - B_N(t) \right\|_{L^2([0,1])} \right\} = 0 .
\end{equation}
Both statements (\ref{eq:Bconv1}, \ref{eq:Bconv2}) can be extended to
arbitrary spectral representation of Brownian motion.  Note also that
these statements are applicable pointwise, e.g.,
\begin{equation}
\label{eq:Bconv3}
\lim\limits_{N\to\infty} \E\left\{ \left| B(t) - B_N(t) \right| \right\} = 0 
\end{equation}
for any $t\in [0,1]$.

\subsection{Nowhere differentiability}

Since Brownian motion is a sum of hat functions it is easy to believe
that $B(t)$ it not differentiable almost everywhere.  One can prove a
much stronger statement that $B(t)$ is nowhere differentiable with
probability one.  The proof goes along the same lines as for
convergence (the argument follows the proof from \cite{Morters}).

We are going to show that almost surely for all $t$ at least one of
the two limits,
\begin{equation*}
\begin{split}
\overline B'(t) & = \limsup \frac{B(t+h)-B(t)}{h},  \\  
\underline B'(t) & =\liminf \frac{B(t+h)-B(t)}{h}. \\
\end{split}
\end{equation*}
is infinite.  This obviously implies that $B(t)$ is almost surely
nowhere differentiable.  Note that at local extrema, one of these
limits can be finite, so it is not true that {\it both} of them are
always infinite.

Let us assume that there is $t_0$ such that both limits are finite at
$t_0$.  This implies that there is a random finite constant $M$ such
that
\begin{equation}
\label{eq:increment}
\left|\frac{B(t_0+h)-B(t_0)}{h} \right|<M	
\end{equation}
for all $h$.  For a given scale $n$, let $k$ be such that $t_0$ is
between dyadic points $t_{n,k-1}$ and $t_{n,k}$, where $t_{n,k} \equiv
2^{-n} k$.  The triangle inequality implies that for any $j$
\begin{equation*}
\begin{split}
& |B(t_{n,k+j}) - B(t_{n,k+j-1})| < |B(t_{n,k+j}) - B(t_0)| \\
& + |B(t_{n,k+j-1}) - B(t_0)| < M(2j+1)2^{-n}.  \\
\end{split}
\end{equation*}

Let $E_{nk}$ be the event that this inequality holds for $j=1,2,3$.
Since increments are independent normal variables, one gets
$\P\{E_{nk}\}\le c (2^{-n/2})^3$, where $c$ is a constant.  The
probability that these inequalities hold for {\it some} $k$ from $0$
to $2^n-1$ is then bounded by $2^n (c 2^{-3n/2}) = c 2^{-n/2}$.  The
sum of these probabilities over $n$ is finite, hence by Borel-Cantelli
lemma, with probability one only finitely many of them will occur.  On
the other hand the assumed inequality (\ref{eq:increment}) implies
that infinitely many of $E_{nk}$ will occur.  This yields the
contradiction and proves that (\ref{eq:increment}) cannot be true.

\section{The 1/3-trick to estimate the distance}
\label{sec:trick_proof}

Although the 1/3-trick is classical in analysis, we provide some
explanations which may be instructive for non-experts.

The trick is based on a very simple result.  Let $I_{nk} = [k2^{-n},
(k+1)2^{-n})$ and its boundary $\partial I_{nk} = \{ k2^{-n},
(k+1)2^{-n}\}$.  For any real $x$ and any integer $n$, there exist two
intervals $I_{nk}$ and $I_{nk'}$ of the same length $2^{-n}$ that $x$
belongs to $I_{nk}$ and $x'=x+1/3$ belongs to $I_{nk'}$.  If the
distance from $x$ to the boundary of $I_{nk}$ is smaller than
$2^{-n}/6$, then the distance from $x'$ to the boundary of $I_{nk'}$
is larger than $2^{-n}/6$, and vice-versa.  In other words, the point
$x$ and the shifted point $x'$ cannot be simultaneously close to the
interval endpoints.

\begin{figure}
\begin{center}
\includegraphics[width=60mm]{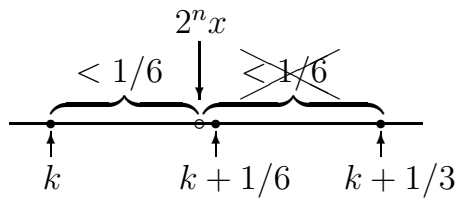}  
\end{center}
\caption{
Illustration for the proof of the $1/3$-trick.  In this example,
$\hat{k} = \hat{k'} + k_0 = k$ and $n$ is even. }
\label{fig:trick}
\end{figure}

Suppose the opposite is true, so that 
\begin{equation*}
|2^n x - \hat{k}| < 1/6 ,  \qquad   |2^n(x+1/3) - \hat{k}'| < 1/6 ,
\end{equation*}
where the integers $\hat{k} = k + [2(2^n x -k)]$ and $\hat{k}' = k' +
[2(2^n x' -k')]$ denote the closest endpoint to $x$ and $x'$,
respectively.  Since $2^n/3 = k_0 + (-1)^n/3$ with an integer $k_0$,
the point $2^n x$ should be simultaneously within the distance $1/6$
to $\hat{k}$ and $\hat{k}' + k_0 + (-1)^n/3$ that is impossible since
the distance between these points is larger than $1/3$
(Fig.~\ref{fig:trick}):
\begin{equation*}
\begin{split}
\frac13 & \leq |(\hat{k}' + k_0 + (-1)^n/3) - \hat{k}| \leq |2^n x - \hat{k}| \\
& + |2^n x - (\hat{k}' + k_0 + (-1)^n/3)| < 1/3. \\
\end{split}
\end{equation*}

Using this simple result, one can prove the estimate for the distance
from a given point $x$ to the set of boundary points $\{y_m\}$.  Let
$I_{nk}$ be the largest interval containing $x$ and not containing any
boundary point $y_m$.  Similarly, for the shifted point $x'=x+1/3$,
let $I_{n'k'}$ be the largest interval containing $x'$ and not
containing any shifted boundary point $y_m' = y_m + 1/3$.  Then the
distance from $x$ to the set of boundary points $\{y_m\}$ is larger
than $2^{-\max\{n,n'\}}/6$.

Suppose that $n\geq n'$.  Assume that the statement is false so there
exists a boundary point $y\in\{y_m\}$ such that $|x-y|<2^{-n}/6$.
Suppose that $y<x$ (the opposite case is similar).  Since $x\in
I_{nk}$ and $y\notin I_{nk}$, both points $x$ and $y$ should be close
to the endpoint $k2^{-n}$:
\begin{equation*}
|x - k 2^{-n}|<2^{-n}/6 ,   \hskip 5mm  |y - k 2^{-n}|<2^{-n}/6 .
\end{equation*}
The shifted boundary point $y'$ belongs to some interval $I_{nj}$.
According to the previous result, the second inequality implies that
the shifted point $y'$ cannot be close to the endpoints of $I_{nj}$:
\begin{equation*}
2^{-n}(j+1/6) \leq y'\leq 2^{-n}(j+5/6) .
\end{equation*}
Since $|x'-y'| = |x-y| < 2^{-n}/6$ and $y<x$, then $y'<x'<y' + 2^{-n}/6$,
hence
\begin{equation*}
2^{-n}(j+1/6) \leq x'< 2^{-n}(j+1) ,
\end{equation*}
i.e., the point $x'$ belongs to the same interval $I_{nj}$ as $y'$.
At the same time, $x'$ belongs to $I_{n'k'}$ which is larger than
$I_{nj}$ due to $n\geq n'$.  The dyadic structure implies that $I_{nj}
\subset I_{n'k'}$ so that $y'$ should belong to $I_{n'k'}$ as well.
But this is in contradiction with the initial assumption about
$I_{n'k'}$.


\end{document}